\documentclass{jfm}

\usepackage[utf8]{inputenc}
\usepackage{graphicx}
\usepackage{placeins}
\usepackage{psfrag}
\usepackage{newtxtext}
\usepackage{newtxmath}
\usepackage{natbib}
\usepackage{datetime}
\usepackage{amsmath}
\usepackage{amsfonts}
\usepackage{amssymb}
\usepackage{commath}
\usepackage{color}
\usepackage{xcolor}
\usepackage{comment}
\usepackage[colorlinks=true, citecolor=black]{hyperref}
\usepackage{mathtools}
\usepackage{soul,tikz}
\usepackage{cancel}
\usepackage{empheq}
\usepackage{nccmath}
\usepackage{tcolorbox}
\usepackage{siunitx}
\usepackage{float}
\usepackage{booktabs}
\usepackage{enumitem}
\usepackage[normalem]{ulem}
\hypersetup{
    colorlinks = true,
    urlcolor   = blue,
    citecolor  = black,
}

\newcommand{\RomanNumeralCaps}[1]

\newcommand{\avertime}[1]{\overline{#1}}
\newcommand{\averzed}[1]{\left\langle #1 \right\rangle}
\newcommand{\avertimezed}[1]{\left[#1\right]}

\newcommand{\noopsort}[1]{}

\usepackage{tikz}
\usetikzlibrary{patterns}
\usepackage{caption}
\usepackage{subcaption}

\title{Turbulent boundary layers altered by passively rotating discs}

\author{Max W. Knoop\aff{1} \corresp{\email{m.w.knoop@tudelft.nl}} \and Pierre Ricco\aff{2}}

\affiliation{
\aff{1} Faculty of Aerospace Engineering, Delft University of Technology, Delft, 2629HS, The Netherlands
\aff{2} School of Mechanical, Aerospace and Civil Engineering, The University of Sheffield, Sheffield, S1 3JD, UK
}

\begin{document}
\maketitle

\begin{abstract}
Turbulent boundary layers characterised by friction Reynolds numbers in the range $Re_{\tau} = 880 - 1460$ and flowing over flush-mounted passively rotating discs are investigated in a wind tunnel with the purpose of reducing the skin-friction drag. The test surface is composed of thirty-two rotating discs arranged in a staggered configuration and supported by bearings mounted in cylindrical cavities. As the discs are half covered by thin rigid plates, a steady rotation of the discs is sustained via the asymmetric distribution of the wall-shear stress exerted by the wall turbulence on the exposed halves of the discs. Direct force measurements reveal that the drag increases with respect to a flat-plate case because of the flow interaction with the disc housings and the covering plates. The effect of the disc motion is isolated and a 3\% drag reduction is measured with respect to the flow over stationary discs. The skin-friction identity by \cite{Elnahhas_Johnson_2022} (\emph{J. Fluid Mech.}, vol. 940, 2022), extended herein to include the disc-flow effects, is utilised for the first time to analyse experimental data. This direct slip effect, quantified by using the measured disc angular velocities in the Elnahhas-Johnson identity, is negligible. Measurements obtained by particle image velocimetry disclose that a roughness mean-flow effect occurs between adjacent discs because of the clearance gaps around the discs and that a downwash secondary flow exists near the covering plates, analogous to flows over streamwise-elongated rectangular roughness elements. This downwash velocity is streamwise modulated because of the spanwise disc motion and alters the wall-normal transport term in the Elnahhas-Johnson identity, thus reducing the drag locally. 
The velocity covariances are enhanced in the proximity of adjacent discs, in line with the roughness effect in that region. The Reynolds-shear-stress term in the Elnahhas-Johnson identity is thereby globally enhanced.
Nonetheless, the near-wall one-point and two-point covariances of the velocity fluctuations and the Reynolds shear stresses are reduced in magnitude and spatial coherence. This result is the first experimental evidence confirming that the attenuation of the Reynolds shear stresses is the central drag-reducing mechanism behind the rotating-disc flow. 

\end{abstract}


%=====================
\section{Introduction}
Wall-bounded turbulence control aimed at reducing skin-friction drag has been widely investigated, motivated by the potential to obtain a net energy saving. Various wall-based active control strategies have been considered, such as reactive wall-normal blowing \citep{choi1994active,abbassi2017skin,dacome2024opposition}, uniform spanwise wall oscillations \citep{jung-mangiavacchi-akhavan-1992, choi2002near,quadrio-ricco-2004} and streamwise travelling waves of spanwise wall velocity \citep{quadrio_streamwise-travelling_2009,quadrio_laminar_2011}. Spanwise wall forcing has been found to achieve drag-reduction margins exceeding 50\% by the periodic interaction of the spanwise velocity profile with near-wall turbulence \citep{quadrio_laminar_2011,Touber_near-wall_2012,agostini_turbulence_2015,knoop2025response}.

Wall-based flush-mounted rotating discs were first proposed as actuators to control wall-bounded turbulent flows by \citet{keefe1997normal}. Motivated by Keefe's idea, Ricco and co-workers investigated a turbulent channel flow over rotating discs using direct numerical simulations (DNS). For arrays of discs rotating in alternate sense along the streamwise direction, \citet{Ricco2013disks} reported a maximum drag reduction of 23\% and a maximum net power saving of 10.5\%, computed by taking into account the theoretical power required to actuate the discs against viscous forces. Boundary layers flowing over rotating discs were investigated experimentally by \citet{Klewicki2003Laminar} in the laminar regime and by \citet{Kempaiah2022Large-scale} in the turbulent regime.

Variants of the original rotating-disc method studied by \citet{Ricco2013disks} have led to comparable or improved performance, i.e. drag-reduction margins as large as 26\% \citep{Wise2014Oscilating,Wise2014SpinningConfig,olivucci2019turbulent}. Using extensions of the \citet*{fukagata2002contribution} identity, \citet{Ricco2013disks} and \citet{olivucci2019turbulent} demonstrated that the drag-reduction mechanism is associated with an overall attenuation of the Reynolds shear stresses, produced by the spanwise-shear effect on the near-wall turbulent structures and by the wall-normal shift of the flow induced by the von-K\'arm\'an rotating boundary layer \citep{karman1921laminare}.

Although active control techniques can achieve large drag-reduction margins, the input power required to run a real-world mechanical system for flow control often exceeds the theoretical input power, thus outweighing the drag-reduction savings. In experimental implementations of the travelling-wave spanwise forcing technique, the actual input power typically exceeds the power savings by three orders of magnitude \citep{auteri2010experimental,gatti2015experimental,bird2018experimental}. Researchers have therefore devoted their attention to passive methods, for which no power input is needed to achieve the drag-reduction effect. One of these methods is a passively rotating disc, first introduced in the experimental study of \citet{Koch2013Drag}. The disc was positioned flush on a flat wall and mounted on a bearing located in an internal housing. Half of the disc was covered by a thin metal plate. The passive motion of the disc was sustained by the torque exerted by the surface-integrated wall-shear stress of the turbulent boundary layer flowing on the exposed half-disc. In fully developed conditions, this torque was balanced by a resistive torque that could be split into three components: one created by the bearing, one exerted by the internal flow between the disc and its housing, and one given by the flow between the disc and the covering plate. \citet{Koch2013Drag} did not measure the drag, but utilised the measurement of the disc motion and an empirical formula for the skin-friction coefficient of wall turbulence to estimate a drag reduction of 17\% on the exposed half-disc. Further details of this study can be found in \citet{koch-kozulovic-2014} and in the PhD thesis of \citet{koch2014reduzierung}.

The first DNS study of wall turbulence altered by passively rotating discs was conducted by \cite{Olivucci2021reduction}. They investigated how the turbulence was modified as it flowed over a fully-exposed disc and half-covered discs arranged periodically along the flat walls of a channel. The Navier-Stokes equations were coupled to the dynamical equations of each disc by integrating the turbulent wall-shear stress over the disc surfaces and by modelling the resistive torques of the bearing and of the internal housing flows. Under the action of turbulence, the fully-exposed disc oscillated without mean rotation. No drag reduction was computed in that case because of the small oscillation amplitude of the disc motion. The half-discs instead spun at a constant rate and an overall maximum drag reduction of 5.6\% was achieved. \cite{Olivucci2021reduction} investigated the influence of the cavity height, the disc diameter and the model of the resistive torques. They reported two mechanisms responsible for the skin-friction drag reduction: the direct influence of the streamwise wall motion on the wall-shear stress and the indirect attenuation of the Reynolds shear stresses caused by the disc rotation.

%=================================
\subsection{Overview of the study}
To the best of the authors' knowledge, the method of passively rotating discs for turbulent skin-friction reduction has only been investigated by \cite{Koch2013Drag} and \cite{Olivucci2021reduction}. Motivated by this limited attention, an experimental study of the passive-disc method was carried out in a wind tunnel at the Delft University of Technology. The test surface was composed of thirty-two discs arranged in a staggered configuration. Measurements included the overall drag exerted by the turbulent boundary layer on the test surface, the disc rotation and the flow statistics over the disc surface. 

The skin-friction identity discovered by \citet{Elnahhas_Johnson_2022} (EJ) for standard turbulent boundary layers was extended to our passive-disc case. Our study marks the first time that this identity is used to investigate a three-dimensional flow and a controlled wall-bounded flow. It is also the first time that the terms in the identity are computed using experimental data.
New terms in the extended EJ identity pertain to (i) the direct-slip effect of the disc motion on the drag reduction, (ii) the three-dimensional flow over the surface, and (iii) the wall-transpiration through the radial clearances between the discs and the fixed portion of the wall. Our extension of the EJ identity is also relevant for other studies involving a wall slip velocity, such as flows over passively moving belts \citep{bechert-etal-1996} and over superhydrophobic surfaces \citep{hervet2003flow,rastegari2019drag}.

We report the impact of the complex geometry of the test surface, which included the disc-housing components and the plates that partially covered the discs, all features that were absent in the channel-flow simulations of \citet{Olivucci2021reduction}. Although an overall drag increase occurred due to the geometry of the test surface, the disc motion reduced the drag with respect to flow over stationary discs. The direct-slip contribution to the overall drag was small, the clearance gaps around the disc produced a roughness effect, and the Reynolds shear stresses were attenuated near the disc surface and enhanced in the bulk of the flow.

The experimental procedures are presented in \S\ref{sec:method} and the EJ identity is extended to the rotating-disc case in \S\ref{sec:EJ}. The modifications of the drag components caused by the disc rotation and the test-surface geometry are discussed in \S\ref{sec:integralDragmeasurements}. In \S\ref{sec:slipDrag}, the direct influence of the streamwise slip on the drag reduction is analysed via the corresponding EJ term. The modifications of the mean flow and the second-order turbulence statistics are presented in \S\ref{sec:baseFlowTopology} and \S\ref{sec:hoStatisticsTopology}. The other EJ terms are evaluated in \S\ref{sec:CfBudget}. A summary and concluding remarks are found in \S\ref{sec:conclusions}. 

%================================
\section{Experimental procedures}
\label{sec:method}

%=================================
\subsection{Experimental facility}
\label{sec:setup}

\begin{figure}
    \centering
    \includegraphics{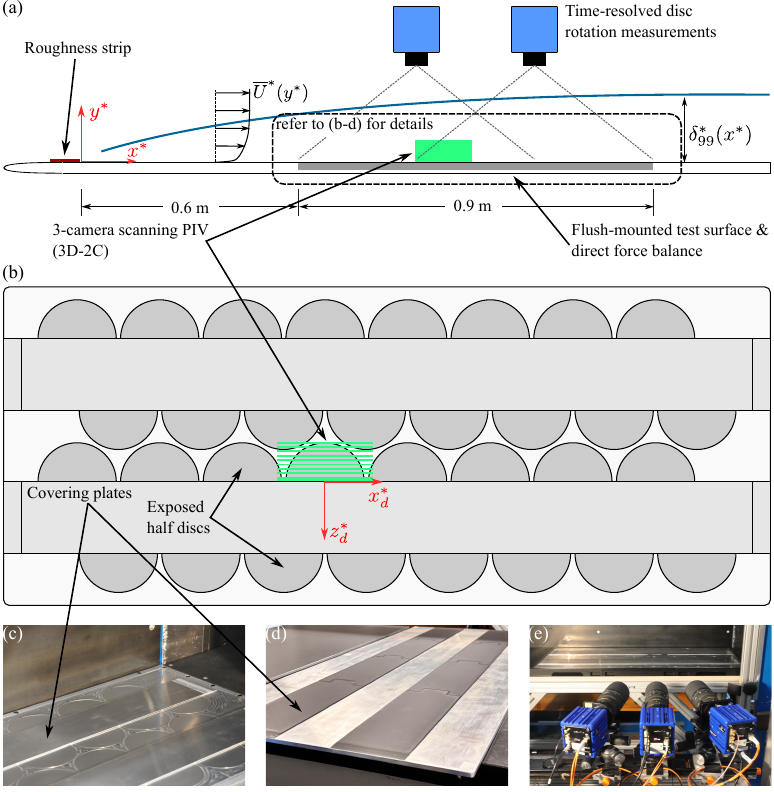}
    \caption{(a) Side-view schematic of the experimental setup. (b) Top-view schematic of the rotating-disc surface and the 10 scanning-PIV planes arranged along the span (PIV stands for particle image velocimetry). The photographs show (c) the rotating-disc surface near the leading edge of the model, (d) the test surface featuring the covering plates only and (e) the 3-camera PIV setup. A video of the rotating discs is available in the supplementary material.}
    \label{fig:expSetup}
\end{figure}

Figure~\ref{fig:expSetup} presents schematics and photographs of the experimental setup. The experiments were conducted at the Delft University of Technology in an open-return wind tunnel, characterised by a cross-section of $0.4 \times0.4$ m$^2$ and a free-stream turbulence intensity of 0.7\% \citep{carrasco2024experimental}. As shown in figure~\ref{fig:expSetup}(a), a boundary layer developed from the elliptical leading edge of the flat-plate test section and was tripped to the turbulent regime by a carborundum roughness strip (24-grit) located downstream of the leading edge. Downstream of the roughness strip and upstream of the test surfaces, the turbulent boundary layer (TBL) had a streamwise development length of $0.6$\:m. 

The origin of the global coordinate system is located at the downstream end of the roughness strip and along the centreline of the tunnel. This coordinate system is defined by the streamwise ($x^*$), wall-normal ($y^*$) and spanwise ($z^*$) directions. The corresponding velocity components are $u^*$, $v^*$ and $w^*$, time is denoted by $t^*$ and pressure is denoted by $p^*$. Dimensional quantities are herein indicated by the superscript *. The test surfaces were located in the streamwise range $x^* = 0.6 - 1.5$\:m. 

A reference free-stream velocity was computed near the leading edge of the test surfaces ($ x^*$ = 0.632\:m) using the particle image velocimetry (PIV) measurements (refer to \S\ref{sec:blChar_viscousScaling}).
$U_{\infty,ref}^*$ varied in the range $18.5-33.8$\:m/s.

%=====================================================
\subsection{Rotating-disc and reference test surfaces}
\label{sec:discModel}
\begin{figure}
    \centering
    \includegraphics{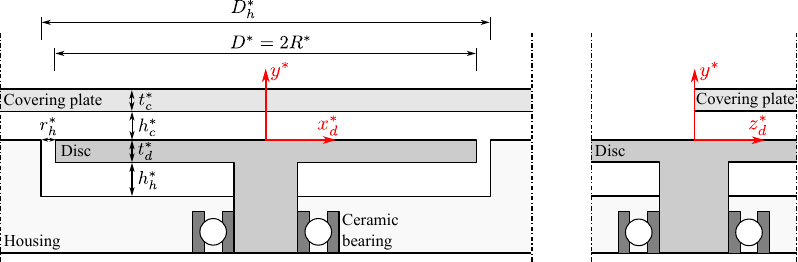}
    \caption{Schematic diagrams of a disc and its housing. The relevant dimensions are given in table~\ref{tab:geometry}. The distances are not to scale. (a) Side view in the $x_d^* - y^*$ plane and (b) cross-section view in the $y^* - z_d^*$ plane.}
    \label{fig:discGeometry}
\end{figure}

The test surface with the passively rotating discs was dimensioned according to the standard geometry for the direct force balance designed and built at the Delft University of Technology \citep{Nesselrooij_development_2022}. The test surface had a thickness of 5\:mm and dimensions of 881.3\:mm along $x^*$ and 366.3\:mm along $z^*$. As shown in figure~\ref{fig:expSetup}(b), it included thirty-two discs arranged in a staggered configuration in four rows along $x^*$. The diameter of the discs was $D^* = 90$\:mm and the centre-to-centre distance between discs was 95\:mm.
Figure~\ref{fig:discGeometry} shows how a disc was mounted flush with the wall and installed in its housing. The dimensions of the disc geometry are given in table~\ref{tab:geometry}. As shown in figure~\ref{fig:expSetup}(b), another coordinate system, denoted by the subscript $d$, is located at the centre of the disc where detailed measurements for the flow statistics were conducted. 

Each disc was attached to its housing using a ceramic bearing (SMB 687 Si$_3$N$_4$). Slots at the bottom of the disc housings were used to immobilise the discs in order to study the influence of the rotation on the drag. A video of the discs in motion is available in the supplementary material. As further shown in the photograph of figure~\ref{fig:expSetup}(c), the discs were half-covered by two thin flat covering plates that extended along the whole streamwise length of the test surface and formed streamwise-aligned steps along the centrelines of the discs. To minimise form drag, the leading and trailing edges of the plates had a triangular taper with a 10:1 aspect ratio.

Two other test surfaces were used as references to quantify the influence of the disc motion on the flow. The first surface was a flat plate on which an uncontrolled TBL developed. The second surface, shown in figure~\ref{fig:expSetup}(d), had the purpose of quantifying the influence of the covering plates. This surface consisted of two solid ridges, but it excluded the discs, the internal housings, the housing gaps around the discs, and the gaps between the covering plates and the bottom surface. Its height was 1.5~mm and the geometry was the same as that of the rotating-disc surface, i.e. having triangular leading and trailing edges with a 10:1 aspect-ratio taper. The ridges were 3D printed using fused deposition modelling with a 0.1 mm layer height and bonded to a flat plate.

%------------
\begin{table}
\setlength{\tabcolsep}{8pt}
\centering
\begin{tabular}{ccccccc}
$D^*$ & $D_h^*$ & $r_h^*$ & $h_h^*$ & $t_d^*$ & $h_c^*$ & $t_c^*$ \\[3pt]
90 & 91 & 0.5 & 1 & 1 & 0.5 & 1\\
\end{tabular}
\caption{Dimensions of the disc system in mm. The nomenclature corresponds to the geometry in figure~\ref{fig:discGeometry}.}
\label{tab:geometry}
\end{table}

%=====================================
\subsection{Direct force measurements} 
\label{sec:dragmsm}
A floating element force balance was used to measure the drag force exerted by a TBL on the test surfaces \citep{Nesselrooij_development_2022}. The balance measurements have been extensively validated in the standard flat-plate configuration. The balance was specifically designed for high-accuracy drag differences required in flow-control studies. The measurements of the drag differences relative to a reference surface are within $\pm0.5\%$ at a 95\% confidence level. This methodology has been used for studies of passive drag-reduction techniques \citep{van2023experimental,hartog2024turbulent, carrasco2024experimental}.

The integral force $F^*$ exerted on a test surface is a combination of the surface-integrated wall-shear stress and the form drag resulting from the geometrical features of the test surface. $F^*$ was computed by correcting the measured raw force for the sensor drift and for the pressure forces on the leading and trailing edges of the floating element \citep[refer to][]{vanNesselrooij2016drag}. 
One measurement consisted of a set of eight equally spaced free-stream velocities in the range $U_{\infty,ref}^* = 18.5-33.8$\:m/s.
For each free-stream velocity, the raw force was acquired at 10~kHz over a 20-s period. Once the free-stream velocity settled to a new value, a 20-s break was included for the discs to reach a steady rotation. The measurements were repeated five times, alternating between the flat plate and the test surface with discs. Further details regarding the drag-measurement procedure are found in \citet{Nesselrooij_development_2022}.

The drag coefficient is defined as:
\begin{equation}
    C_D = \frac{2F^*}{\rho^* U_{\infty,ref}^{*2}S^*},
\end{equation}
where $S^*$ is the surface area of the flat plate and $\rho^*$ is the density of the fluid. The percentage difference in drag with respect to the flat-plate reference surface is:
\begin{equation}
    \label{eq:DeltaCD}
    \Delta C_D (\%) = 100 \cdot \left(\frac{C_D}{C_{D,0}} - 1\right),
\end{equation}
where the subscript `0' hereafter denotes quantities pertaining to the flat-plate TBL. Negative values of $\Delta C_D$ correspond to drag reduction, whereas positive values correspond to drag increase. The uncertainty was quantified by the variance of the five repeated measurements. Uncertainty propagation was applied to obtain the standard error and 95\% confidence bounds were computed using the Student's t-distribution to account for the small sample size of the five measurements.

%=============================================
\subsection{Measurements of the disc rotation}
\label{sec:rotmsm}
Time-resolved image sequences were acquired to measure the angular velocity of the discs and the distribution of the wall velocity. As depicted in figure~\ref{fig:expSetup}(a), two Photron Fastcam Mini AX100 high-speed cameras ($1024 \times1024$ pixels, 20\:\textmu m pixel size, 12-bit) were used to capture the motion of the discs from a top view. The cameras were mounted with 20-mm Nikkor-AF objectives at an f/2.8 aperture and captured a field of view of $0.7 \times 0.7$\:m$^2$ and $0.5 \times 0.5$\:m$^2$. 10\,000 individual images were acquired for each free-stream condition. The image acquisition frequency was varied in the range $f^*=1.6-3$~kHz to account for the increase in the angular velocity of the discs with the free-stream velocity. The number of captured rotations varied in the range $N=30-80$, depending on the angular velocity of the disc.

\begin{figure}
    \centering
    \includegraphics{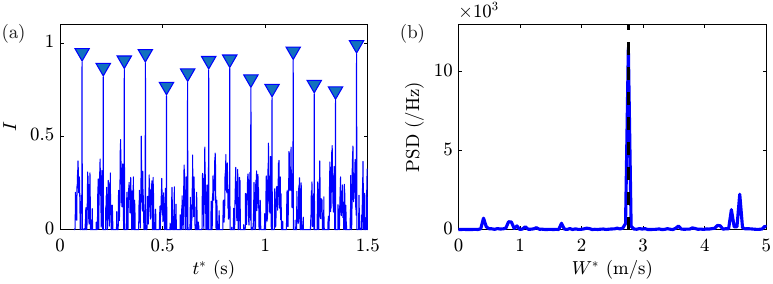}
    \caption{(a) Time series of the indicator for the measurement of the disc angular velocity. The triangles denote the result of the peak detection algorithm. (b) Power spectral density of the indicator and time-averaged disc tip velocity, denoted by the vertical dashed line.}
    \label{fig:discRotMethod}
\end{figure}
 
Each disc was marked at its circumference to measure the disc velocity. The marker was tracked via a time series of the minimum intensity in a $4\times 10$ pixel window in the azimuthal and radial directions. As shown in figure~\ref{fig:discRotMethod}(a), the data were rescaled to obtain an indicator between 0 and 1 so that a marker-passing corresponded to a peak of about unity. A peak detection algorithm was used to identify the time-separation $\Delta T^*$ between consecutive marker passings, represented by the blue triangles in figure~\ref{fig:discRotMethod}(a). Inconsistencies in the detection algorithm were removed by filtering outliers of the $\Delta T^*$ vector based on the median absolute deviation. The disc angular velocity was computed as $\Omega^* = 2\pi/\Delta T^*$ and the disc tip velocity was computed as $W^* = D^*\Omega^*/2$. As shown in figure~\ref{fig:discRotMethod}(b), the time-averaged disc tip velocity $\avertime{W}^*$, indicated by the dashed line, was consistent with the peak of the power spectral density (PSD) of the indicator. The disc-rotation rate was steady for all discs, as evidenced by the equidistant marker spacings in figure~\ref{fig:discRotMethod}(a) and the narrow PSD peak in figure~\ref{fig:discRotMethod}(b).

Two aspects were considered in the computation of the uncertainty associated with the angular position of the marker and, correspondingly, $W^*$: (i) the angular displacement of the marker between time steps, i.e. the product of the time separation between individual images ($\Delta T^* = 1/f^*$) and the disc tip velocity, and (ii) the angular size of the interrogation window (4\:pixels), based on the measurement resolution $M^* = 1.8$ and $2.2$ pixels/mm for the two cameras. The two uncertainty components were of the same order of magnitude. By assuming two uncorrelated components of uncertainty, the uncertainty of a single measurement of the disc tip velocity is defined as $\epsilon_W^*=\epsilon_\theta^*/\Delta T^*,$ where
\begin{equation}
    \label{eq:unc_x}
    \epsilon_\theta^* 
    = 
    \sqrt{\left(\frac{W^*}{f^*}\right)^2+\left(\frac{4}{M^*}\right)^2}.
\end{equation}
Assuming $N$ independent samples, the time-averaged uncertainty of the disc tip velocity is $\epsilon_{\avertime{W}}^* = \avertime{\epsilon_W^*}/\sqrt{N}$. It was computed to be small, i.e. $\epsilon_{\avertime{W}}^*$=0.003-0.007\:m/s or 0.13-0.15\%.

%======================================
\subsection{Particle image velocimetry over a passively rotating disc}
\label{sec:PIV}
Measurements were conducted using three-camera PIV in $x^* - y^*$ planes and scanning along the $z^*$ direction. The objective was to obtain volumetric two-component flow statistics (3D-2C) over an exposed half-disc. 
The evolution of the TBL was captured over the fourth disc, located in the second row and centred at $x^* = 0.97$\:m, as shown in figure~\ref{fig:expSetup}(b). 
The TBL on the flat plate was also measured at a single $z^*$ location. 
Flow quantities denoted by the subscript $d$ refer to the location $x_d^*=-D^*/2$, regardless of whether the flat-plate surface or the rotating-disc surface was used. The free-stream velocity in the disc coordinate system was $U_{\infty,d}^* = 18.8$\:m/s. Details on how $U_\infty^*$ was computed are given in \S\ref{sec:blChar_viscousScaling}.

Three digital LaVision sCMOS cameras ($2560 \times 2160$ pixels, 6.5 \textmu m pixel size, 16-bit) were mounted with 200-mm Nikkor-AF objectives at an f/4 aperture, capturing a combined field-of-view of size $94 \times 25$\:mm$^2$ in each $x^* - y^*$ plane. Figure~\ref{fig:expSetup}(e) shows the camera configuration, where the upstream and downstream cameras were oriented at an angle of $\pm10^\circ$ around the $y$-axis to create a 3-mm overlap between the camera views. These cameras were mounted with Scheimpflug adapters to ensure uniform focus. A pinhole model, calibrated using millimetre paper, provided a transfer function between the image and the physical coordinates. The laser sheet was moved in the spanwise direction by a Zaber LRQ150HL traverse in the range $-45\:\text{mm}$$\leq$$ z_d^*$$\leq$$-2.5\:\text{mm}$. The first increment was 2.5\:mm and subsequent planes were spaced by 5~mm, for a total of 10 measurement planes. In each measurement plane, 2000 image pairs were acquired at a frequency of 15~Hz, providing uncorrelated snapshots separated by approximately $55\delta_{99}^*/U_{\infty,ref}^*$ boundary-layer turnover times, where the boundary-layer thickness $\delta_{99}^*$ is defined in \S\ref{sec:blChar_viscousScaling}. The time-separation between the individual image pairs was $\mathrm{d}t^* = 14$\:\textmu s, yielding 20-pixel particle displacements in the free stream.

\begin{sloppypar}
Instantaneous velocity vectors were obtained using a multi-pass cross-correlation algorithm and an adaptive interrogation-window deformation method \citep{scarano2000advances}. The image pre-processing procedure was as follows. A sliding time filter was first applied by subtracting the minimum over nine images. Spatial minimum subtraction and normalisation were then employed within a 5-pixel kernel and Gaussian smoothing with a 2-pixel radius. The cross-correlation was evaluated in elliptical interrogation windows of $96 \times 24$ pixels, with a 75\% overlap, ensuring higher resolution in the wall-normal direction. The viscous resolution of an interrogation window was $1.2 \times 0.3$\:mm$^2$ or $60 \times 15\:\delta_{\nu,0}^{*2}$ in reference viscous units, where $\delta_{\nu,0}^*$ is the viscous length scale defined in \S\ref{sec:blChar_viscousScaling}. Velocity outliers were detected and removed using a universal detection algorithm \citep{westerweel2005universal}, applying a median filter within a $5 \times 5$-pixel window.
\end{sloppypar}

%==============================================================
\subsection{Viscous scaling and boundary-layer characteristics} 
\label{sec:blChar_viscousScaling}

Viscous scaling was adopted using two reference friction velocities, $u_{\tau,0}^*$ and $u_{\tau,d}^*$, and the corresponding viscous length scales, $\delta_{\nu,0}^*$ and $\delta_{\nu,d}^*$, where $\nu^*$ is the kinematic viscosity of the fluid. The surface-averaged reference friction velocity $u_{\tau,0}^*$ was used for the results pertaining to global variables (refer to \S\ref{sec:integralDragmeasurements} and \S\ref{sec:slipDrag}). The friction velocity $u_{\tau,d}^*$ was used for the results in the disc coordinate system pertaining to the PIV data (refer to \S\ref{sec:baseFlowTopology}, \S\ref{sec:hoStatisticsTopology} and \S\ref{sec:CfBudget}). As the streamwise wall-shear stress was not measured along $x^*$, the surface-averaged friction velocity was defined as $u_{\tau,0}^* = \sqrt{F_0^*/(\rho^* S^*)}$, where $F_0^*$ is the force exerted by the wall turbulence on the reference flat plate. 
The reference skin-friction coefficient is $C_{f,0} = 2(u_{\tau,0}^*/U_{\infty,ref}^*)^2$.
In the disc coordinate system, $u_{\tau,d}^*$ was obtained by fitting the time-averaged streamwise velocity of the flat-plate TBL to the composite profile of \citet{chauhan2009criteria} with log-layer constants $\kappa = 0.384$ and $B = 4.17$. 
The corresponding skin-friction coefficient is  $C_{f,d} = 2(u_{\tau,d}^*/U_{\infty,d}^*)^2$.
Quantities scaled in viscous units are denoted by the superscript +.

Mean streamwise-velocity profiles over the flat plate were obtained by PIV using one camera (300 image pairs) and by streamwise averaging over $\pm 2$\:mm at $x^* = 632$ and 1472\:mm, which are the leading-edge and the trailing-edge locations of the flat plate. The free-stream velocity $U_{\infty}^*(x)$ was computed by time, wall-normal and spanwise averaging the PIV measurements of $u^*$ across $y^*/\delta_{99}^* \in [1.25,1.4]$ and $z_d^*/D^* \in [-0.5,0]$. Spanwise and wall-normal pressure gradient effects in the free stream could be neglected since the spanwise variations of the free-stream velocity remained below $0.15\%$ and the wall-normal variations remained below $0.05\%$. The local boundary-layer thickness $\delta_{99}^*(x)$ was computed as the wall-normal height where the time-averaged streamwise velocity was $0.99U_{\infty}^*$. It was denoted by $\delta_{99,1}^*$ and $\delta_{99,2}^*$ at the leading-edge and trailing-edge locations.
At these locations, the momentum thicknesses $\theta_{1}^*$ and $\theta_{2}^*$, defined in \eqref{eq:theta}, and the respective momentum-thickness Reynolds numbers $Re_\theta = U_{\infty}^*\theta^*/\nu^*$ were computed.
Since $u_{\tau,0}^*$ is an integral quantity, an average friction Reynolds number $Re_{\tau,avg} = u_{\tau,0}^*(\delta_{99,1}^*+\delta_{99,2}^*)/2\nu^*$ was computed using the averaged boundary-layer thickness as a length scale.
Table~\ref{tab:BL_chars} presents these flow parameters for the drag-balance measurements. As expected, the boundary-layer and momentum thicknesses grew along the streamwise direction and decreased as $U_{\infty,ref}^*$ increased.

The flow parameters for PIV over the rotating disc for $U_{\infty,d}^* = 18.8$\:m/s are shown in table~\ref{tab:conditionsPIV}. The friction velocity for the TBL over the discs was larger due to roughness effects and contains a viscous and a pressure-drag component, as discussed in \S\ref{sec:MeanFlowAndEJ-1}. 
Table~\ref{tab:discScaling} presents the outer-scaled and viscous-scaled disc diameter, the covering-plate step height, and the radial clearance gap size.

\begin{table}
\centering
\begin{tabular}{ccccccccccc}
$U_{\infty,ref}^*$  & $C_{f,0}\times 10^{3}$ & $u_{\tau,0}^*$  & $\delta_{\nu,0}^*$  & $\theta_{1}^* $ & $\theta^*_{2}$ & $\delta_{99,1}^*$  & $\delta_{99,2}^*$  & $Re_{\theta,1}$ & $Re_{\theta,2}$ & $Re_{\tau,avg}$  \\
(m/s) &  & (m/s) & (\textmu m) & (mm)&  (mm) & (mm) & (mm) &  &  \\
\midrule
18.5 & 2.97 & 0.714 & 20.5 & 1.63 & 2.63 & 13.1 & 22.8 & 1980 & 3350 & 880 \\
20.7 & 2.92 & 0.791 & 18.5 & 1.60 & 2.59 & 13.0 & 22.5 & 2160 & 3680 & 960  \\
22.9 & 2.87 & 0.868 & 16.8 & 1.55 & 2.59 & 12.8 & 22.1 & 2330 & 4070 & 1030 \\
25.1 & 2.83 & 0.944 & 15.5 & 1.54 & 2.53 & 12.5 & 22.4 & 2530 & 4360 & 1130 \\
27.3 & 2.80 & 1.020 & 14.3 & 1.51 & 2.57 & 12.4 & 22.5 & 2690 & 4820 & 1220  \\
29.4 & 2.77 & 1.094 & 13.4 & 1.52 & 2.51 & 12.3 & 22.1 & 2930 & 5090 & 1290 \\
31.6 & 2.74 & 1.169 & 12.5 & 1.46 & 2.59 & 11.9 & 22.2 & 3010 & 5640 & 1360 \\
33.8 & 2.72 & 1.245 & 11.8 & 1.47 & 2.58 & 12.5 & 21.9 & 3250 & 6010 & 1460 \\
\end{tabular}
\caption{Experimental conditions and flat-plate TBL parameters for the drag-balance measurements.}
\label{tab:BL_chars}
\end{table}

\begin{table}
    \centering
    \begin{tabular}{cccccccccc}
     & $U_{\infty,d}^*$ &$C_{f,d}\times 10^{3}$ & $u_{\tau,d}^*$  & $\delta_{\nu,d}^*$  & $\theta^*_d$ & $\delta_{99,d}^*$ & $Re_{\theta,d}$ & $Re_{\tau,d}$ \\
     &  (m/s)& & (m/s) & (\textmu m) &  (mm) &  (mm) &  &  \\   
     \midrule
    TBL over flat plate  & 18.8 & 3.15 & 0.747 & 19.5 & 2.46 & 17.8 & 2640 & 900 \\
    TBL over discs       & 18.8 & 4.15 & 0.858 & 17.3 & 2.65 & 18.5 & 2840 & 1070
    \end{tabular}
    \caption{TBL parameters for the PIV measurements at the leading edge of the disc ($x_d^* = -D^*/2$). Refer to figure~\ref{fig:expSetup}(b) for the measurement location.}
    \label{tab:conditionsPIV}
\end{table}

\begin{table}
\setlength{\tabcolsep}{7pt}
    \centering
    \begin{tabular}{ccc|ccc}
         $D^*/\delta_{99}^*$ & $(t_c^*+h_c^*)/\delta_{99}^*$ & $r_h^*/\delta_{99}^*$ & $D^+$ & $t_c^+ + h_c^+$ & $r_h^+$\\
         5 & 0.084 & 0.028 & 4610 & 76.8 & 25.6
    \end{tabular}
    \caption{Disc-geometry quantities scaled in outer and viscous units, based on the flat-plate TBL conditions given in table~\ref{tab:conditionsPIV}.}
    \label{tab:discScaling}
\end{table}

%============================================================
\section{Elnahhas-Johnson identity for the passive-disc flow}
\label{sec:EJ}
In \S\ref{sec:derivation}, we extend the \citet{Elnahhas_Johnson_2022} identity for the skin-friction coefficient of a canonical TBL to the case of passively rotating discs. 
% The extended EJ identity provides a decomposition of the skin-friction coefficient altered by the discs as the sum of (i) a term that isolates the laminar skin-friction coefficient over a flat fixed plate, (ii) a term related to the Reynolds stresses, (iii) a term given by the streamwise pressure gradient, (iv) terms related to the non-parallel convective transport and (v) new terms that are produced by the disc rotation. These new terms are given by (1) the streamwise slip caused by the disc motion, (2) the spanwise time-averaged flow and (3) the transpiration through the clearance between the disc and the fixed portion of the wall. 
The derivation of the EJ identity relies on the assumption that the TBL evolves on a flat surface. Therefore, the influence of the unevenness of the covering plates and of the flow in the internal housing is not taken into account. The terms of the identity are computed by using the spatially-resolved wall velocity and the PIV data, as detailed in \S\ref{sec:EJPIV}.

\subsection{Derivation of the extended Elnahhas-Johnson identity}
\label{sec:derivation}
Quantities are scaled as follows: $\{u,v,w\}=\{u^*,v^*,w^*\}/ U_{\infty,ref}^*$, $\{x,y,z\}=\{x^*,y^*,z^*\}/D^*$, $t = t^* U_{\infty,ref}^*/D^*$ and $p = p^*/(\rho^* U_{\infty,ref}^{*2})$. Although $D=1$, this value is not assigned to $D$ for clarity.
The following averaging operators of a quantity $q$ are defined:
\begin{equation}
\label{eq:avertime}
\mbox{time averaging}:
\avertime{q} = \frac{1}{T} \int_0^{T} q \ \mathrm{d }t,
\end{equation}
\begin{equation}
\mbox{spanwise averaging}:
\averzed{q} = \frac{1}{z_2-z_1} \int_{z_1}^{z_2} q \ \mathrm{d}z,
\end{equation}
\begin{equation}
\label{eq:avertimezed}
\mbox{spanwise and time averaging}:
\avertimezed{q} = \frac{1}{T(z_2-z_1)} \int_0^{T} \int_{z_1}^{z_2} q \ \mathrm{d}z \ \mathrm{d}t = \averzed{\avertime{q}},
\end{equation}
where $z_{d,1}=-D/2$ and $z_{d,2}=0$, as shown in figure \ref{fig:expSetup}(b). The covering plates are therefore excluded from the domain of integration. Fluctuating quantities are defined with respect to the local time-averaged flow as $q'(x,y,z,t) = q(x,y,z,t) - \avertime{q}(x,y,z)$.

The skin-friction coefficient is:
\begin{equation}
\label{eq:cf}
C_f(x) = \left.\dfrac{2 \nu}{U_\infty(x)^2}\frac{\p \avertimezed{u}}{\p y}\right|_{y=0},
\end{equation}
where $\nu = \nu^*/(U_{\infty,ref}^* D^*)$.

To derive the extended EJ identity, the streamwise momentum deficit equation is first obtained. As discussed in \S\ref{sec:PIV}, the free-stream flow can be assumed to be uniform in the wall-normal and spanwise directions. The time-averaged free-stream flow is described by the streamwise momentum equation
\begin{equation}
\label{eq:free-stream}
U_\infty \dfrac{\mathrm{d} U_\infty}{\mathrm{d} x} + \dfrac{\mathrm{d} P_\infty}{\mathrm{d} x}=0.
\end{equation}
By using the time-averaged continuity equation, equation \eqref{eq:free-stream} and the time-averaged streamwise momentum equation, the streamwise momentum deficit equation is found,
\begin{equation}
\label{eq:time-avg-x-mom-deficit}
\begin{aligned}
&
\dfrac{\p (U_\infty-\avertime{u})\avertime{u}}{\p x}
+\dfrac{\p (U_\infty-\avertime{u})\avertime{v}}{\p y}
+\dfrac{\p (U_\infty-\avertime{u}) \avertime{w}}{\p z}
+(U_\infty-\avertime{u})\dfrac{\mathrm{d} U_\infty}{\mathrm{d} x}
=
\\ &
- \nu \dfrac{\p^2 \avertime{u}}{\p y^{2}}
+ \dfrac{\p \avertime{u' v'}}{\p y}
-
\underbrace{
\left(
\dfrac{\p (P_\infty-\avertime{p})}{\p x}
+ \nu \dfrac{\p^2 \avertime{u}}{\p x^{2}}
- \dfrac{\p \avertime{u' u'}}{\p x}
\right)
}_{I_{x}}
-
\underbrace{
\left(
\nu \dfrac{\p^2 \avertime{u}}{\p z^{2}}
- \dfrac{\p \avertime{u' w'}}{\p z}
\right)
}_{I_{z}},
\end{aligned}
\end{equation}
where the terms $I_{x}$ and $I_{z}$ are negligible for a flat-plate TBL according to high-Reynolds-number boundary-layer theory. Equation \eqref{eq:time-avg-x-mom-deficit} is averaged along the spanwise direction to find
\begin{equation}
\label{eq:mom-deficit-eq}
\begin{aligned}
&
 \dfrac{\p \averzed{(U_\infty-\avertime{u})\avertime{u}}}{\p x}
+\dfrac{\p \averzed{(U_\infty-\avertime{u})\avertime{v}}}{\p y}
+W_d(x_d,y)
+(U_\infty-\avertimezed{u})\dfrac{\mathrm{d} U_\infty}{\mathrm{d} x}
\\ &
+ \nu \dfrac{\p^2 \avertimezed{u}}{\p y^{2}}
- \dfrac{\p \avertimezed{u' v'}}{\p y}
+\averzed{I_{x}}
+\averzed{I_{z}}
=0,
\end{aligned}
\end{equation}
where
\begin{equation}
\label{eq:Wd}
\begin{aligned}
W_d(x_d,y)=
\dfrac{2}{D}
\bigg(\big(U_\infty-\avertime{u}(z_{d,2}) \big) \avertime{w}(z_{d,2})-\big(U_\infty-\avertime{u}\left(z_{d,1}\right)\big) \avertime{w}\left(z_{d,1}\right)
\bigg),
\end{aligned}
\end{equation}
\begin{equation}
\label{eq:<Ix>}
\averzed{I_{x}}=
\dfrac{\p \left(P_\infty-\avertimezed{p}\right)}{\p x} + \nu \dfrac{\p^2 \avertimezed{u}}{\p x^{2}}-\dfrac{\p \avertimezed{u' u'}}{\p x},
\end{equation}
\begin{equation}
\label{eq:<Iz>}
\averzed{I_{z}}=
\frac{2}{D}\left( \nu \left.\dfrac{\p \avertime{u}}{\p z}\right|_{z_{d,2}} - \nu \left.\dfrac{\p \avertime{u}}{\p z}\right|_{z_{d,1}}-\avertime{u' w'}(z_{d,2})+ \avertime{u' w'}(z_{d,1})\right).
\end{equation}
To obtain the skin-friction identity, equation \eqref{eq:mom-deficit-eq} is multiplied by $y-l(x)$ (where $l(x)$ is a length to be determined) and integrated in the wall-normal direction from 0 to $\infty$.

The integrated wall-normal diffusion term is
\begin{equation}
\label{eq:ejViscous}
\begin{aligned}
%&
\nu \int_0^\infty \left(y-l(x)\right) \frac{\p^2\avertimezed{u}}{\p y^2} \mathrm{d}y =
%\\ &
-l(x) U_\infty(x)^2
\left(
    -\dfrac{C_f(x)}{2}
    +\dfrac{\nu}{l(x) U_\infty(x)}
    -\dfrac{U_w \nu}{l(x) U_\infty(x)^2}
\right),
\end{aligned}
\end{equation}
where $U_w(x) = \avertimezed{u}(x,y=0)$. The new third term on the right-hand side of \eqref{eq:ejViscous} captures the direct influence of the wall motion on the skin-friction coefficient.

The integrated wall-normal transport term is
\begin{equation}
\label{eq:EJ-ytransport}
\begin{aligned}
&
\int_0^\infty (y-l(x)) \dfrac{\p \averzed{(U_\infty-\avertime{u})\avertime{v}}}{\p y} \mathrm{d}y =
-l(x) U_\infty(x)^2
\left(
    -\averzed{\left(1-\dfrac{\avertime{u}(y=0)}{U_\infty(x)}\right)\dfrac{\avertime{v}(y=0)}{U_\infty(x)}}
    +\dfrac{\theta_v(x)}{l(x)}
\right),
\end{aligned}
\end{equation}
where $\avertime{u}(y=0) = \Omega^*z_d^*$ and
\begin{equation}
\theta_v
=
\int_0^\infty
\averzed{
\left(1 - \dfrac{\avertime{u}}{U_\infty(x)}\right)\dfrac{\avertime{v}}{U_\infty(x)}
}
\mathrm{d}y.
\end{equation}
The new first term in \eqref{eq:EJ-ytransport} quantifies the effect of wall-normal transpiration through the clearance between the disc and the fixed portion of the wall.

The integrated spanwise transport term is
\begin{equation}
\int_0^\infty (y-l(x)) W_d(x,y) \mathrm{d}y =
-l(x)U_\infty(x)^2 \int_0^\infty \left(1-\dfrac{y}{l(x)}\right) \dfrac{W_d(x,y)}{U_\infty(x)^2} \mathrm{d}y.
\end{equation}
The following integral thicknesses are defined:

\begin{equation}
\label{eq:theta}
\theta(x) = \int_0^\infty \averzed{\left(1 - \dfrac{\avertime{u}}{U_\infty(x)}\right)\dfrac{\avertime{u}}{U_\infty(x)}}\mathrm{d}y,
\end{equation}
\begin{equation}
\delta_l(x)
=
\int_0^\infty
\left(1-\dfrac{y}{l(x)}\right)
\left(1 - \dfrac{\avertimezed{u}}{U_\infty(x)}\right)
\mathrm{d}y,
\end{equation}
\begin{equation}
\theta_l(x) = \int_0^\infty \left(1-\frac{y}{l(x)}\right) \averzed{\left(1 - \dfrac{\avertime{u}}{U_\infty(x)}\right)\dfrac{\avertime{u}}{U_\infty(x)}}\mathrm{d}y.
\end{equation}

Dividing the weighted integration of \eqref{eq:mom-deficit-eq} by $l(x) U_\infty(x)^2$ and solving for $C_f/2$, the extended EJ identity is obtained:
\begin{equation}
\label{eq:EJ}
\begin{aligned}
\dfrac{C_f(x)}{2}=
&
    \underbrace{\dfrac{1}{Re_l(x)}}_{\mathcal{L}}
    \underbrace{-\dfrac{1}{Re_l(x)}\dfrac{U_w}{U_\infty(x)}}_{\mathcal{S}}
    +\underbrace{\int_0^\infty \dfrac{-\avertimezed{u' v'}}{U_\infty(x)^2 l(x)} \mathrm{d}y}_{\mathcal{T}}
    \\ &
    +\underbrace{\dfrac{\mathrm{d}\theta_l(x)}{\mathrm{d} x}
    -\dfrac{\theta(x)-\theta_l(x)}{l(x)} \dfrac{\mathrm{d}l(x)}{\mathrm{d}x}}_{\mathcal{N}_x}
    +\underbrace{\dfrac{\theta_v(x)}{l(x)}}_{\mathcal{N}_y}
    \underbrace{-\averzed{\left(1-\dfrac{\avertime{u}(y=0)}{U_\infty(x)}\right)\dfrac{\avertime{v}(y=0)}{U_\infty(x)}}}_{\mathcal{V}_w}
    \\ &
    +\underbrace{\int_0^\infty \left(1-\dfrac{y}{l(x)}\right) \dfrac{W_d(x,y)}{U_\infty(x)^2} \mathrm{d}y}_{\mathcal{N}_z}
   +\underbrace{\dfrac{\delta_l(x)+2 \theta_l(x)}{U_\infty(x)}\dfrac{\mathrm{d} U_\infty(x)}{\mathrm{d} x}}_{\mathcal{P}_x}
\\ &
+\underbrace{\int_0^\infty \left(1-\dfrac{y}{l(x)}\right) \dfrac{\averzed{I_{x}}}{U_\infty(x)^2} \mathrm{d}y}_{\mathcal{I}_{x}}
+\underbrace{\int_0^\infty \left(1-\dfrac{y}{l(x)}\right) \dfrac{\averzed{I_{z}}}{U_\infty(x)^2} \mathrm{d}y}_{\mathcal{I}_{z}},
\end{aligned}
\end{equation}
where we use $Re_l(x)=U_\infty(x) l(x)/\nu$ to keep the notation as close as possible to that of the original EJ formulation. To summarise, the effect of the disc motion on the skin-friction coefficient is given by $\mathcal{S}(x)$, the direct change of the skin-friction coefficient due to the wall motion, $\mathcal{V}_w(x)$, the impact of the wall-transpiration through the disc clearance, $\mathcal{N}_z(x)$, the influence of the spanwise flow induced by the disc motion, and $\mathcal{I}_{z}(x)$ due to the new terms \eqref{eq:<Iz>} at the spanwise sides of the integration domain, caused by the spanwise viscous diffusion and the spanwise velocity fluctuations. When $U_w=0$, $W_d(x,y)=0, \avertime{v}(y=0)=0$ and $\mathcal{I}_{z}(x)=0$, identity \eqref{eq:EJ} simplifies to identity (2.22) by EJ. When $\avertimezed{u' v'}=0$ and the wall is stationary and smooth, the boundary layer becomes the Blasius boundary layer and $C_f(x)=2\mathcal{L}$, i.e. the skin-friction coefficient reduces to the laminar value. The terms $\mathcal{N}_x$ and $\mathcal{N}_y$ in \eqref{eq:EJ} quantify the influence of non-parallelism of the turbulent mean flow on the skin-friction coefficient with respect to the laminar flow. They arise from the first two terms in \eqref{eq:mom-deficit-eq}, i.e. the streamwise and wall-normal transports of momentum deficit. For a Blasius boundary layer, these terms sum to zero because they simplify to $-A/Re_l(x)$ and $A/Re_l(x)$, where $A$ is a constant.

%=====================================================================
\subsection{Computation of the terms in the Elnahhas-Johnson identity}
\label{sec:EJPIV}
The data from the PIV measurements over the fourth downstream half-disc in the second row, depicted in figure~\ref{fig:expSetup}(b), were used to compute the terms in the EJ identity \eqref{eq:EJ}. The turbulence statistics in the individual PIV planes were spanwise and time averaged using \eqref{eq:avertimezed} in the range $-0.5 \leq z_d \leq 0$. A weighted sum was used for the spanwise average because the ten planes of measurements were not equidistant. The velocity $U_w$ in $\mathcal{S}$ was directly obtained from the disc rotation measurements, as described in \S\ref{sec:rotmsm}. As suggested by EJ, the momentum thickness $\theta(x)$ was chosen as the reference scale to compute $l(x) = 4.54\theta(x)$, because the results based on $\theta$ are less sensitive to the exact upstream evolution and transition history of the boundary layer. This choice was appropriate in our case because the TBL might have been influenced by the TBL tripping (refer to \S\ref{sec:setup}).

The wall-normal integrals in \eqref{eq:EJ} were computed using trapezoidal integration in the range $15$$\leq y^+$$\leq 1200$ ($0.003$$ \leq$$y/\delta_{99}$$\leq$$1.4$), where the upper bound extended to the edge of the PIV field-of-view. Gradients were computed using a second-order accurate central-difference scheme. Turbulent fluctuations with length scales smaller than the PIV interrogation windows were not resolved due to the spatial averaging in windows of size $60\times 15$\:$\delta_{\nu,0}^{*2}$ ($x^*\times y^*$). This spatial averaging caused an attenuation of the wall-normal velocity variances and the Reynolds stresses for $y^+$$<$$200$ and, in turn, impacted the computation of $\mathcal{T}$. This attenuation effect has been amply documented in several studies when comparing TBL statistics measured via PIV with those obtained via DNS \citep{lee2016validating}.

\begin{figure}
    \centering
    \includegraphics{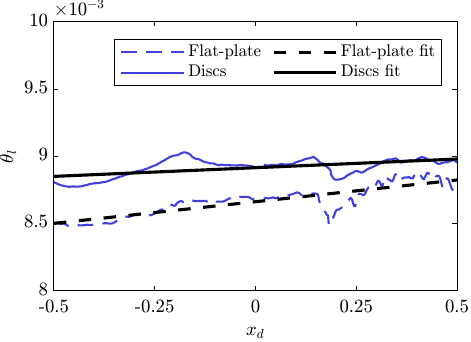}
    \caption{Streamwise variation of $\theta_l$ (blue lines) and first-order polynomial fits (black lines) for TBL over the flat plate (dashed lines) and the rotating discs (solid lines).}
    \label{fig:thetalFit}
\end{figure}

Figure~\ref{fig:thetalFit} depicts the streamwise variation of $\theta_l$. The local peaks are due to the stitching procedure applied to the overlapping frames of the 3-camera PIV measurements. Because of these variations, a first-order polynomial least-squares regression of $\theta_l$ was applied in the range -0.5$\leq$$x_d$$\leq$0.5, denoted by the black line in figure~\ref{fig:thetalFit}. The quantity $\mathrm{d}\theta_l/\mathrm{d}x$, needed for the computation of $\mathcal{N}_x$, was calculated using the regression line.
The computation of $\mathcal{N}_y$ carried a large uncertainty because it involves the integral term $\theta_v$ that contains the wall-normal velocity $\avertime{v}$. The uncertainty in the measurement of $\avertime{v}$ was larger than that of $\avertime{u}$ because of the large anisotropy between the two flow components \citep{adrian1997dynamic,wilson2013uncertainty}.
The spanwise mean flow, required for the computation of $\mathcal{N}_z$, was not directly available from our PIV measurements. We first calculated $\p \avertime{w}/\p z$ by computing $\p \avertime{u}/\p x$ and $\p \avertime{v}/\p y$ through finite-difference schemes and by using the time-averaged continuity equation. We then estimated $\avertime{w}$ by integrating $\p \avertime{w}/\p z$ along $z$. $\mathcal{N}_z$ could, however, not be computed reliably due to the large uncertainty in determining $\avertime{w}$. The primary reason for this uncertainty was the discrete spacing of the PIV planes, which led to an accumulating error when integrating the continuity terms along the spanwise direction.

The pressure-gradient term $\mathcal{P}_x$ \eqref{eq:EJ} required the computation of $\mathrm{d}U_\infty/\mathrm{d}x$. Similar to $\theta_l(x)$ in figure~\ref{fig:thetalFit}, $U_\infty(x)$ shows a trend that could be approximated by a first-order polynomial fit. A maximum deviation of 0.2\% from its polynomial fit was found.

The streamwise-diffusion term involving $\avertime{u}$ and the convection term involving $\avertime{u' u'}$ in $I_{x}$ were verified to have a contribution of less than $0.1\%$ of $C_{f,0}$ for the flows over the flat plate and the discs. The second and the third terms of $I_{x}$ in \eqref{eq:time-avg-x-mom-deficit} are therefore negligible. 
The streamwise pressure gradient term in $I_{x}$, although negligible in classical flat-plate TBL, could play a role in the balance of the flows over the disc surfaces and between discs. Its measurement or numerical computation remain an important point for future research. This component could be the only non-negligible term of $\mathcal{I}_x$ in \eqref{eq:EJ}.

The spanwise-diffusion term involving $\p\avertime{u}/\p z$, part of $I_{z}$ in \eqref{eq:time-avg-x-mom-deficit}, is also negligible (computed to contribute less than $0.1\%$ of $C_{f,0}$) since the wall-parallel streamwise and spanwise gradients over a disc can be assumed to be comparable. The terms involving $\avertime{u' w'}$ in \eqref{eq:time-avg-x-mom-deficit} were not available from the PIV measurements, and therefore their contribution to $\mathcal{I}_z$ could not be computed.

%================
\section{Results}
\label{sec:results}
The drag characteristics of the flow induced by the passively rotating discs are presented in \S\ref{sec:integralDragmeasurements}, focusing on the effects of the disc rotation, the covering plates and the disc housings. The influence on the drag of the direct-slip component $\mathcal{S}$ in the extended EJ identity \eqref{eq:EJ} is discussed in \S\ref{sec:slipDrag}.
The modifications by the discs of the mean-flow and second-order flow statistics are discussed in \S\ref{sec:baseFlowTopology} and \S\ref{sec:hoStatisticsTopology}, respectively. The relation of these flow modifications to the EJ skin-friction budget is studied further in \S\ref{sec:CfBudget}.

%=========================================
\subsection{Integral drag characteristics}
\label{sec:integralDragmeasurements}
The direct force measurements, described in \S\ref{sec:dragmsm}, are used to evaluate the drag characteristics of the flow produced by the passively rotating discs. The following drag changes, related to the three test configurations discussed in \S\ref{sec:discModel}, are considered: \\
\begin{enumerate}[label=(\roman*), leftmargin=*]
    \item \ $\Delta C_{D,RD}$, produced by the passively rotating discs when they are free to rotate under the torque of the wall turbulence. It evaluates the influence of the test surface with respect to standard flat-plate geometry when all the features of the drag-reduction system are present, i.e., discs in motion, gaps around discs, disc housings, and covering plates.\\
    \item \ $\Delta C_{D,NRD}$, produced by the non-rotating discs, i.e., the configuration is the same as in (i), but the discs are mechanically fixed. It evaluates the influence of the stationary discs, disc gaps and housings, and covering plates. \\
    \item \ $\Delta C_{D,cover}$, produced by the covering-plate geometry shown in figure~\ref{fig:expSetup}(d), when used in lieu of the disc surface employed for (i) and (ii). It evaluates the influence of the covering plates (without discs, disc gaps and housings, and gaps below the covering plates) with respect to the flat-plate case. \\
\end{enumerate} 

The following drag changes can be computed from the terms (i), (ii), (iii) in order to assess further the effects produced by the disc motion and the geometry of the test surface: \\
\begin{itemize}
    \item[$\bullet$] \ $\Delta C_{D,motion}=\Delta C_{D,RD} - \Delta C_{D,NRD}$ evaluates the influence of the disc motion on the change in drag.\\
    \item[$\bullet$] \ $\Delta C_{D,housing}=\Delta C_{D,NRD} - \Delta C_{D,cover}$ evaluates the influence of the disc gaps and housings on the change in drag. \\
\end{itemize}

\begin{figure}
    \centering
    \includegraphics[width=13.5cm]{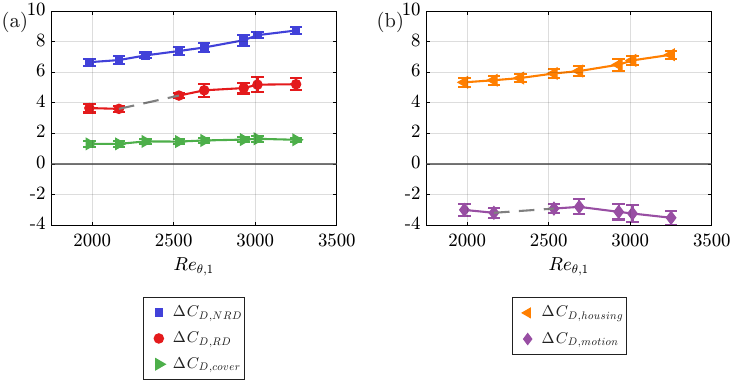}
    \caption{Percentage difference in drag coefficients with respect to the reference flat-plate case as a function of $Re_{\theta,1}$. Error bars denote the 95\% confidence interval (refer to \S\ref{sec:dragmsm}).}
    \label{fig:DeltaCD}
\end{figure}

Figure~\ref{fig:DeltaCD}(a) shows that $\Delta C_{D,RD}$ and $\Delta C_{D,NRD}$ are positive, proving that, regardless of the disc motion, the presence of the test surface is associated with an overall drag increase. The measurement of the rotating case at $Re_{\theta,1} = 2330$ ($Re_{\tau,avg} = 1030$) is omitted because a resonance of the balance system occurred, resulting in spurious force readings due to signal clipping. $\Delta C_{D,RD}= 3.5-5\%$ is lower than $\Delta C_{D,NRD}= 6.5-9\%$, i.e. the disc motion generates a reduction of drag with respect to the flow configuration with stationary discs. The drag reduction produced by the disc motion is quantified by $\Delta C_{D,motion} =  -3\%$, shown by the purple diamonds in figure~\ref{fig:DeltaCD}(b). It does not vary significantly with the Reynolds number in the range $Re_{\theta,1} = 1980-2350$ ($Re_{\tau,avg} = 880 - 1460$) because $\Delta C_{D,RD}$ and $\Delta C_{D,NRD}$ monotonically increase with $Re_{\theta,1}$ at approximately the same rate.
Although both the surface with rotating discs and the surface with fixed discs produce a drag increase with respect to the standard flat-plate case, our result is the first experimental evidence that a rotating disc motion reduces the drag with respect to the same surface with non-rotating discs.

Figure~\ref{fig:DeltaCD}(a) also shows that the covering plates enhance the drag by $\Delta C_{D,cover} = 1.7\%$ with a minimal Reynolds number variation. This increase is likely caused by secondary motions at the sides of the plates and by the friction-drag alteration on the top part of the covering plates, analogous to the flow produced by spanwise-heterogeneous rectangular roughness elements \citep{vanderwel2019instantaneous,medjnoun2020effects}. This result agrees well with the experimental findings of \citet{frohnapfel2024flow}. They reported a $1-2$\% drag increase for a flow over streamwise-elongated rectangular elements of width $1-2\,\delta_{99}^*$, a flow that is analogous to ours. Similar to our case, they also showed that the drag increase was almost independent of the Reynolds number. The comparison of our disc flow with the flow over the rectangular elements further demonstrates that the additional friction induced by the streamwise flow below the covering plates when the discs are stationary is negligible in the overall drag balance.

Figure~\ref{fig:DeltaCD}(b) also reveals that $\Delta C_{D,housing} =  5-7.5\%$ , i.e. the gaps around the discs contribute the most to the increase in drag. This increase is attributed to local separation at the disc gaps.
The surface complexity of the disc housings thus causes a roughness-type flow, as discussed in \S\ref{sec:baseFlowTopology}. The drag increase due to these geometrical features is also responsible for the indirect enhancement of the Reynolds shear stresses, as discussed in \S\ref{sec:hoStatisticsTopology}. The $\Delta C_{D,housing}$ increment with $Re_{\theta,1}$ can be ascribed to the increasing contribution of this roughness effect since the viscous-scaled characteristic roughness length grows as the gap size in viscous units changes in the range $r_h^+= 25-42$ as $Re_{\theta,1}$ varies.

The drag increase due to the geometry of the test surface is, however, of the same order as the drag reduction produced by the disc rotation, highlighting opportunities for optimisation of this flow-control technique.

%=======================================================================
\subsection{Contribution of the streamwise wall slip to the drag change}
\label{sec:slipDrag}

\begin{table}
\centering
\def~{\hphantom{0}}
\begin{tabular}{cccccc|cccccc}
\multicolumn{6}{c}{Present experiment}  & \multicolumn{6}{c}{DNS \citet{Olivucci2021reduction}}    \\
$U_{\infty,ref}^*$ & $Re_{\tau,avg}$ & $D^+$ & $U_s^+$ & $U_{s,d}^+$ & $\Delta C_{D,motion}$ & $U_b^*$ & $Re_\tau$ & $D^+$ & $U_s^+$ & $U_{s,d}^+$ & $\Delta C_D$\\
(m/s) & & & & & (\%) & (m/s) & & & & & (\%)\\
\midrule
18.5    & 880       & 4390  & 0.42      &  1.33  &  $-3.0$       & 0.028       & 180       & 605   & 0.15      & 0.42 & ~$-2.1$    \\
20.7    & 960       & 4870  & 0.47      &  1.49  &  $-3.2$       & 0.028       & 180       & 1210  & 0.07      & 0.42 & ~$-2.1$    \\
22.9    & 1030      & 5360  & 0.50      &  1.58  &  \ \ --       & 0.028       & 180       & 605   & 0.21      & 0.6  & ~$-2.8$    \\
25.1    & 1130      & 5810  & 0.49      &  1.56  &  $-2.9$       & 0.028       & 180       & 605   & 0.72      & 2.06 & $-5.6$    \\
27.3    & 1220      & 6290  & 0.51      &  1.61  &  $-2.8$       & 0.028       & 180       & 605   & 0.37      & 1.05 & $-5.1$    \\
29.4    & 1290      & 6710  & 0.51      &  1.63  &  $-3.1$       &             &           &       &           &                   \\
31.6    & 1360      & 7200  & 0.52      &  1.64  &  $-3.2 $      &             &           &       &           &                   \\
33.8    & 1460      & 7600  & 0.54      &  1.72  &  $-3.5 $      &             &           &       &           &                   \\
\end{tabular}
\caption{Viscous-scaled disc diameter, integral slip-velocities and drag changes for the present rotating-disc experiment and the DNS study of \citet{Olivucci2021reduction}. $U_b^*$ denotes the channel-flow bulk velocity.}
\label{tab:SlipVelocities}
\end{table}

The direct contribution of the disc motion to the drag reduction is estimated. Two integral measures of the streamwise slip are computed. The slip velocity $U_s$ is found by averaging $\avertime{u}(x,y=0,z)$ over the test surface. The slip velocity $U_{s,d}$ is obtained by averaging $\avertime{u}(x,y=0,z)$ over the exposed surfaces of the half-discs. Although the surface-averaged slip effect given by the discs bears some analogy to that produced by the slip-length boundary condition used for flows over superhydrophobic surfaces \citep{min2004effects}, crucial differences are that, in the disc case, the slip effect is large-scale, highly spatially non-uniform and discontinuous because of the disc geometry.

Table~\ref{tab:SlipVelocities} shows the flow conditions and the disc slip velocities of our study and of the channel-flow DNS investigation of \cite{Olivucci2021reduction} at $Re_\tau = 180$. Our disc slip velocities averaged over the whole test surface, $U_s^+ = 0.42-0.54$, are comparable to the two largest $U_s^+$ reported by \cite{Olivucci2021reduction}, although the disc diameters and the Reynolds numbers are different. Our disc slip velocities averaged over the disc surfaces, $U_{s,d}^+=1.33-1.72$, are instead all larger than those computed by \cite{Olivucci2021reduction}, except for the numerical case for $U_{s,d}^+=2.06$.

\begin{figure}
    \centering
    \includegraphics{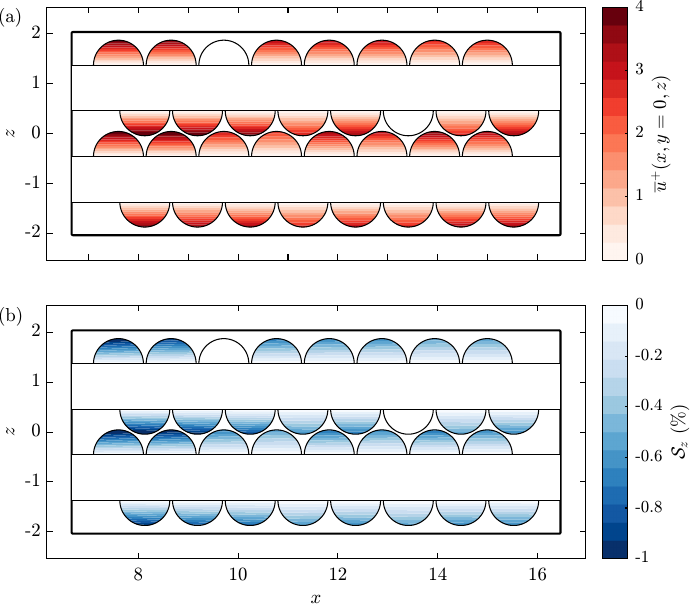}
     \caption{Spatial variations of (a) the streamwise wall velocity at {$Re_{\theta,1}=1980$ ($Re_{\tau,avg} = 880$)} and (b) the direct-slip term $\mathcal{S}_z(x,z)$, defined in \eqref{eq:DeltaCfslipEJ} using $l(x)=4.54\theta_0(x)$, relative to $C_{f,0}/2$.}
    \label{fig:Wp_18ms}
\end{figure}

Figure~\ref{fig:Wp_18ms}(a) shows the time-averaged streamwise wall velocity $\avertime{u}^+(x,y=0,z)$ due to the disc motion at $Re_{\theta,1}=1980$ ($Re_{\tau,avg} = 880$). Because of the disc rigidity, the velocity of each disc varies linearly with the spanwise coordinate and is independent of the streamwise direction. The maximum velocity at the disc tips varies in the range $\avertime{u}^+(x,y=0,z) = 2.5-4.5$ along the test surface. The wall velocity decreases downstream along the upstream half of the test surface ($x < 12$), while it is almost streamwise independent along the downstream half ($x > 12$). As shown in figure~\ref{fig:Wp_18ms}(a), there is a small difference between the time-averaged angular velocities of discs whose tips are at the same $x$ locations because the flows over the discs whose tip is at $z=0$ are more affected by adjacent discs than the flows over the discs farthest away from $z=0$, which are only confined by discs directly downstream and upstream. Two discs were not set in motion by the wall-shear stress, arguably because of manufacturing tolerance and variations of the bearing performance.

We can also estimate the spatial variation of the direct drag modification due to the disc motion by using the $z-$dependent version of the slip term $\mathcal{S}(x)$ in the EJ identity \eqref{eq:EJ}. The $z-$dependent version of $\mathcal{S}(x)$ involves the measured spatial variation of the wall velocity:

\begin{equation}
    \label{eq:DeltaCfslipEJ}
    % \mathcal{S}_z(x,z) = \frac{-2 \ \avertime{u}(x,y=0,z)}{Re_l(x)}.
    \mathcal{S}_z(x,z) = -\frac{1}{Re_l(x)}\frac{\avertime{u}(x,y=0,z)}{U_\infty(x)}.
\end{equation}
The two quantities are related by $\mathcal{S}(x)=\averzed{\mathcal{S}_z(x,z)}$. Figure~\ref{fig:Wp_18ms}(b) shows the spatial variation of $\mathcal{S}_z(x,z)$ as a percentage of $C_{f,0}$ at $Re_{\tau,avg} = 880$. The direct slip causes the wall-shear stress reduction $\mathcal{S}_z(x,z)$ to decay along $x$ more rapidly than $\avertime{u}(x,y=0,z)$ because of the increase of $l(x)$. $\mathcal{S}_z$ reaches a maximum $-1$\% at the disc tips in the upstream region of the test surface, where the local streamwise wall velocity is largest.

\begin{figure}
    \centering
    \includegraphics[width=9cm]{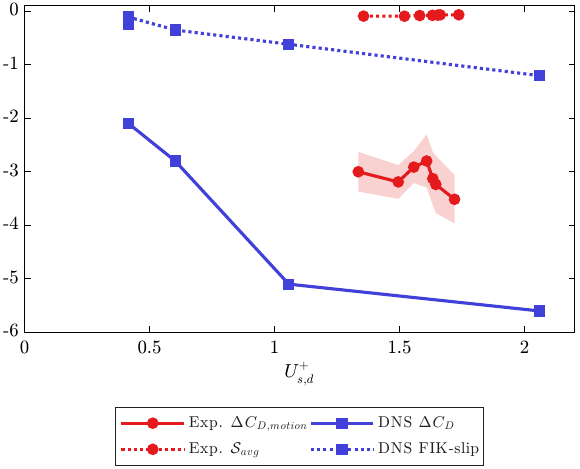}
    \caption{Percentage drag changes as functions of $U_{s,d}^+$ for the present experiments (red circles) and the DNS study of \citet{Olivucci2021reduction} (blue squares). The red-shaded patch indicates the 95\% confidence interval of $\Delta C_{D,motion}$. The integral drag values (solid line) are compared with the direct-slip contribution based on the extended EJ/FIK identities (dotted line). $\mathcal{S}_{avg}$ corresponds to $\mathcal{S}_z$ in figure~\ref{fig:Wp_18ms}(b) averaged over the test surface.}
    \label{fig:DeltaCD_Usp}
\end{figure}

Figure~\ref{fig:DeltaCD_Usp} shows how our measured $\Delta C_{D,motion}$ due to the disc rotation and the DNS drag-reduction values computed by \citet{Olivucci2021reduction} depend on $U_{s,d}^+$. The DNS values follow two drag-reduction regimes. For $U_{s,d}^+<1$, the drag reduces sharply, while, for $U_{s,d}^+>1$, the drag changes much more gradually and saturates to 6\%. Our measured slip velocities belong to the second regime. Our drag-reduction values decrease at a similar rate to the DNS data, albeit within a 95\% confidence interval, denoted by the shaded patch. The data scatter is due to the uncertainty in measuring sub-percentage drag differences. Our measured drag changes depend only marginally on the slip velocity and are lower than the DNS values, i.e. $\Delta C_{D,motion} = -3\%$. Since $\mathcal{S}_z$ only contributes to the overall drag reduction near the disc tips, the average of $\mathcal{S}_z$ over the test surface offers a negligibly small contribution to the overall balance, i.e. $\mathcal{S}_{avg}=-0.1\%$. This result proves that other indirect mechanisms contribute more significantly to the overall change of drag than the streamwise slip. The direct-slip drag reduction in the DNS case, estimated by the extension of the \citet*{fukagata2002contribution} (FIK) identity used by \citet{olivucci2019turbulent,Olivucci2021reduction}, is larger than our experimental $\mathcal{S}_{avg}$, reaching a maximum $-1.2$\%.

In view of our results on the direct slip effect on the drag reduction, the method adopted by \citet{Koch2013Drag} to model the overall drag does not appear to be applicable because it is based solely on the change of skin-friction coefficient due to a streamwise shift of the mean velocity, thus neglecting modifications of the wall turbulence.

%============================================================
\subsection{Statistics of mean velocity over a rotating disc} 
\label{sec:baseFlowTopology}

In this section, the mean-flow modifications are presented and discussed in relation to the drag-reduction effect. All results presented in \S\ref{sec:baseFlowTopology}, \S\ref{sec:hoStatisticsTopology} and \S\ref{sec:CfBudget} pertain to the PIV measurements over a half-disc for $U_{\infty,d}^* = 18.8$\:m/s. Viscous scaling was adopted using the flat-plate TBL parameters given in table~\ref{tab:conditionsPIV}.

%=====================================
\subsubsection{Mean-flow modification} 
\label{sec:mean-flow-modification}

\begin{figure}
    \centering
    \includegraphics{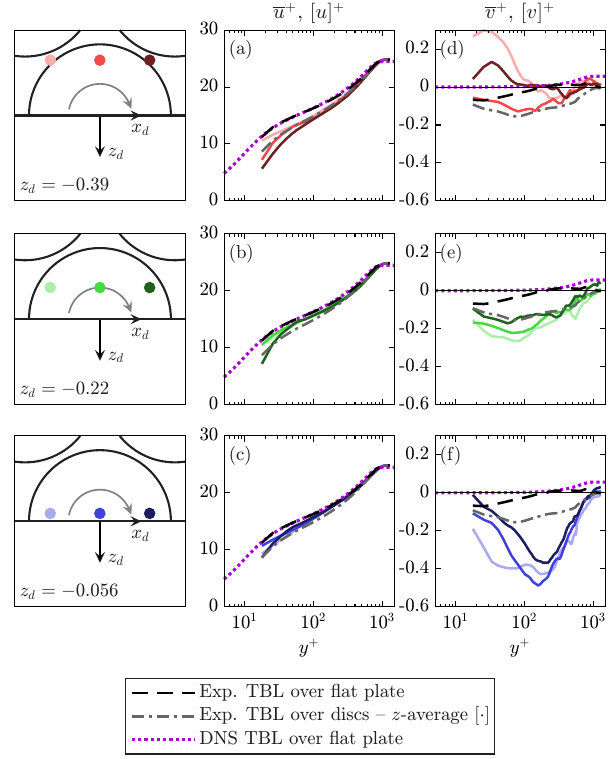}
    \caption{Wall-normal profiles, viscous scaled with $u_{\tau,d}^*$ of the flat plate, of (a-c) mean streamwise velocity and (d-f) mean wall-normal velocity. The black dashed lines denote the flat-plate reference data, the grey dash-dotted lines correspond to the spanwise-averaged value $\avertimezed{\cdot}$ defined in \eqref{eq:avertimezed}, both averaged in the streamwise direction over $-0.5\leq x_d\leq 0.5$, the purple dotted lines show the DNS data of a TBL at $Re_\tau = 830$ by \citet{schlatter2010assessment}. The schematics in the first column indicate the location of the profiles on the disc. The rows correspond to the spanwise location of (a,d) $z_d = -0.39$ in red, (b,e) $z_d = -0.22$ in green and (c,f) $z_d = -0.056$ in blue. The colour gradient (light to dark) denotes the variation of streamwise location $x_d = [-0.35,\ 0,\ 0.35]$.}
    \label{fig:meanStatisticsProfiles}
\end{figure}

Figure~\ref{fig:meanStatisticsProfiles} shows wall-normal profiles of the local mean-flow statistics obtained from the PIV measurements at different locations over the half-disc (the rows correspond to different spanwise locations). The schematics on the left panels of figure~\ref{fig:meanStatisticsProfiles} show the $(x_d,z_d)$ locations of the profiles and the colour gradients denote the streamwise variation. 

\begin{figure}
    \centering
    \includegraphics{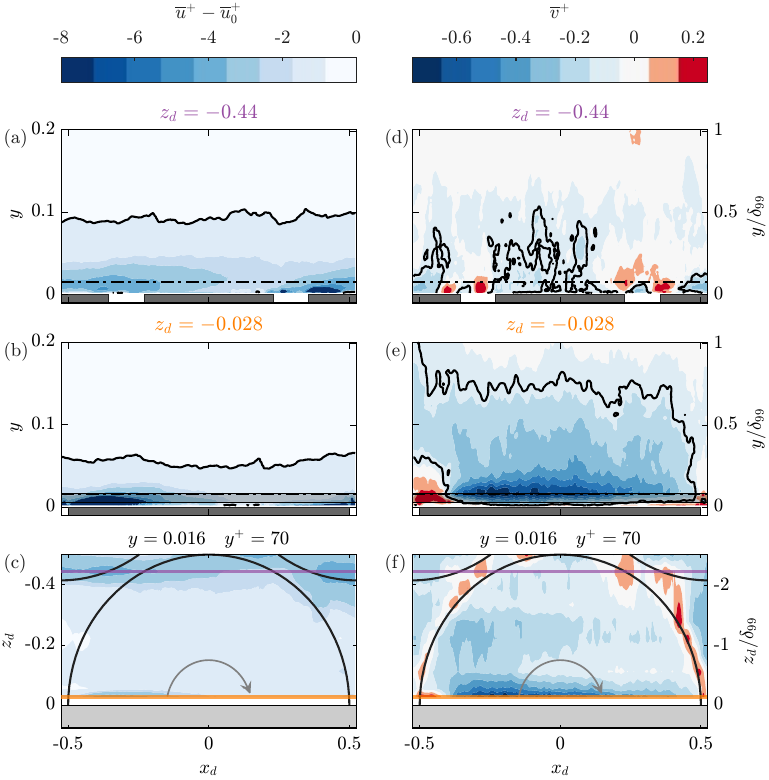}
    \caption{Contours of the disc-induced mean-flow (a-c) streamwise-velocity deficit and (d-f) wall-normal velocity. $x_d-y$ contours at (a,d) $z_d = -0.44$ ($z_d^* = -40$\:mm), (b,e) $z_d = -0.028$ ($z_d^* = -2.5$\:mm), and (c,f) $x_d-z_d$ contours at $y=0.016$ ($y^* = 1.5$\:mm). The wall-normal height of (c,f) is indicated by the dash-dotted lines (a,b,d,e). The coloured lines in (c,f) denote the spanwise location of (a,b,d,e) and match the title colour accordingly. The dark grey regions in (a,b,d,e) represent the cross-section of the rotating discs. The transparent grey regions in (b,e) represent the cross-section of the covering plates. The light grey region in (c,f) indicates the cover. The black contour lines in (a-d) indicate the $y$ location at which the flow modification penetrates and correspond to 5\% of the minimum value within the measurement volume, i.e. $0.05\cdot \min [\avertime{u}^+(x,y,z)-\avertime{u}^+_0(x,y,z)]$ and $0.05\cdot \min \avertime{v}^+(x,y,z)$.}
    \label{fig:meanFlowContours}
\end{figure}

The $\avertime{u}^+$ profiles in figures~\ref{fig:meanStatisticsProfiles}(a-c) indicate that the rotating disc causes a significant streamwise-velocity deficit $\avertime{u}^+-\avertime{u}_0^+$ when compared with the flat-plate case. As shown in figure~\ref{fig:meanStatisticsProfiles}(a), this deficit is largest near the disc tip, in the proximity of the other rotating discs and far from the covering plates. 
% Figure~\ref{fig:meanStatisticsProfiles}(a) shows that, near the disc tip, the deficit is larger at the most downstream location. 
The wall-normal distance to which the deficit extends decreases from $y/\delta_{99} = 0.5$ at $z_d = -0.39$ (figure~\ref{fig:meanStatisticsProfiles}a), to $y/\delta_{99}= 0.3$ at $z_d = -0.056$ (figure~\ref{fig:meanStatisticsProfiles}c). 
The spatial distribution of the deficit is displayed in the contours of figures~\ref{fig:meanFlowContours}(a-c) in side-view $y - z_d$ planes and in top-view $x_d - z_d$ planes. The side view at $z_d = -0.44$ in figure~\ref{fig:meanFlowContours}(a) confirms that the deficit is large near the disc tips and between the two rows of discs. The side view at $z_d = -0.028$ in figure~\ref{fig:meanFlowContours}(b) shows that a large local deficit of $\avertime{u}^+-\avertime{u}_0^+ = -7$ also occurs very near the covering plates and in the upstream part of the disc, i.e. where the disc moves spanwise and away from the plates. The penetration depth of the velocity deficit is indicated by the black contour lines in figures~\ref{fig:meanFlowContours}(a,b), which correspond to $5\%$ of the maximum deficit. In agreement with the results in figures~\ref{fig:meanStatisticsProfiles}(a-c), this wall-normal penetration becomes larger moving in the spanwise direction away from the covering plates. 
% The top view in figure~\ref{fig:meanFlowContours}(c) shows that the spanwise extent where the deficit occurs is much narrower near the covering plates than near the disc tips, i.e. near the purple line $z_d = -0.44$ and between the disc tips.

The profiles of $\avertime{v}^+$ in figures~\ref{fig:meanStatisticsProfiles}(d-f) demonstrate the occurrence of a significant downwash motion (i.e. $\avertime{v}<0$) distributed across the entire disc and extending towards the free stream. This wall-normal velocity is of opposite sign to that pertaining to a canonical TBL and, as indicated by the purple dotted lines, exceeds the value at the edge of the canonical TBL by an order of magnitude \citep[$\avertime{v}^+ = 0.05$;][]{schlatter2010assessment}. The largest downwash motion of up to $\avertime{v}^+ = -0.5$ occurs at $y^+ = 100$ near the covering plates at $z_d = -0.056$. This motion decreases along $z_d$ away from the covering plates, settling to $\avertime{v}^+ = -0.2$ over the disc. The contour plots in figure~\ref{fig:meanFlowContours}(d,e) confirm that the downwash motion is less intense at $z_d = -0.44$ and occurs predominantly near the covering plates, i.e. it is largest at $z_d = -0.028$. At this spanwise location, the downwash motion extends to $y/\delta_{99} = 0.8$, with its peak centred at around $y = 0.016$. This peak location coincides with the top side of the covering plate, where the plate is represented by the transparent grey region in figures~\ref{fig:meanFlowContours}(b,e). The top view at $y = 0.016$ in figure~\ref{fig:meanFlowContours}(f) shows that the peak of the high-magnitude downwash motion forms a streamwise-aligned region with a narrow spanwise extent of $-0.1<z_d<0$ near the covering plates. Local regions of $\avertime{v}^+>0$ also exist. Shown in the top-view contour of figure~\ref{fig:meanFlowContours}(f) and in the profiles of figure~\ref{fig:meanStatisticsProfiles}(d), these regions occur at the disc edges, which may indicate an outflow from the 0.5-mm clearance gaps around the discs.

Although the flow altered by the test-surface geometry is modulated by the disc motion as in the DNS study of \citet{Olivucci2021reduction},
most of the mean-flow modifications reported herein were not found in that study. The covering plates and the other geometrical features of the test surface, all absent in the DNS study, are therefore responsible for the main mean-flow changes, such as the streamwise-velocity deficit near the disc tips and the downwash motion near the covering plates.

%=======================================================================
\subsubsection{Roughness effects produced by the disc gaps and housings}
\label{sec:MeanFlowAndEJ-1}

The large mean-streamwise velocity deficit reported in figures \ref{fig:meanStatisticsProfiles} and \ref{fig:meanFlowContours} is likely caused by a roughness-type effect as the flow evolves over the discs. As shown in figure~\ref{fig:discGeometry}, the flow departs from the hydrodynamically smooth regime because the radial gaps around the discs create local cavity flows, analogous to d-type roughness flows \citep{kadivar2021review}. Although the width of these gaps is physically small, $r_h^* = 0.5$\:mm, it is not negligible when scaled in viscous units, namely $r_h^+ = 25-42$. For flows over rough surfaces, the roughness elements cause a momentum exchange involving the mean streamwise velocity, thereby producing a pressure drag. These effects occur in the inner layer, while an outer-layer similarity is typically retained for $\delta_\nu \ll y \leq \delta_{99}$ \citep{kadivar2021review, chung2021predicting}. 

The $\avertime{u}^+$ profiles and momentum deficit contours in figures~\ref{fig:meanStatisticsProfiles} and \ref{fig:meanFlowContours} are consistent with these roughness-flow characteristics. Near $z_d = 0$, figure~\ref{fig:meanFlowContours}(b) reports that a large momentum deficit occurs at the leading edge of the disc, directly downstream of the disc clearance. Figures~\ref{fig:meanFlowContours}(a,c) instead show that, near the disc tips, the momentum deficit is more spatially distributed. The velocity deficit exists where the interaction between the flow and the surface is particularly complex, characterised by a transverse shear layer between discs rotating in opposing directions and the roughness effect produced by a succession of clearance gaps.

\begin{figure}
    \centering
    \includegraphics{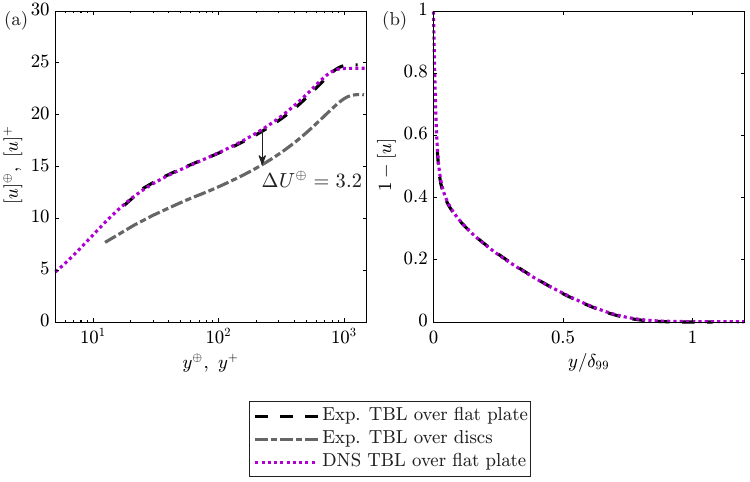}
    \caption{(a) Streamwise-averaged $\avertimezed{u}$ profiles scaled with the actual friction velocity $u_{\tau,d}^*$ and denoted by the superscript $\oplus$. (b) Streamwise-velocity deficit profiles. The line styles correspond to figure~\ref{fig:meanStatisticsProfiles}.}
    \label{fig:roughnessProfiles}
\end{figure}

The stress at the wall, $\tau_{w,r}^*$, which now comprises a viscous component and a pressure-drag component, was determined via a modified composite fit that included a shift of the logarithmic layer ($\Delta U$) as an additional fitting parameter. This procedure is analogous to the Clauser chart method, validated in Appendix A of \citet{knoop2025response}. The computed friction values are reported in table~\ref{tab:conditionsPIV}. While the total wall stress increases due to the pressure-drag component, the viscous component $\mu^* \p \avertime{u}^*/\p y^*|_{y^*=0}$ reduces relative to the flat-plate TBL because of the incurred velocity deficit, where $\mu^*$ is the dynamic viscosity. As shown in figure~\ref{fig:roughnessProfiles}(a), when the $\avertimezed{u^*}$ profiles averaged over the disc surface are scaled by the local $u_{\tau,d}^*=u_{\tau,r}^*=\sqrt{\tau_{w,r}^*/\rho^*}$ over the discs (denoted by superscript $\oplus$), the typical characteristics of a roughness flow emerge. The profiles are self-similar and a downward shift characterised by the roughness function $\Delta U^\oplus = 3.2$ is produced. Outer-layer similarity is also evident from the velocity deficit profiles in figure~\ref{fig:roughnessProfiles}(b). This roughness function belongs to the intermediate roughness regime, with an equivalent sand-grain roughness of $k_s^\oplus = 10$ \citep{Cebeci1977Momentum, kadivar2021review}. The wall stress is increased by $\Delta \tau_{w,r}(\%) = 100\cdot(\tau_{w,r}^{*}/\tau_{w,0}^{*}-1)= 28\%$, which constitutes the leading contribution to $\Delta C_{D,NRD}= 6.5\%$. By multiplying this local wall-stress increase by the coverage ratio of the discs ($0.3$), an estimate for its integral effect is $\Delta \tau_{w,r} = 8.5\%$. This change compares satisfactorily with $\Delta C_{D,NRD}$, considering that the influence of the disc motion and of the spatial development is excluded.

%=====================================================================
\subsubsection{Secondary-flow effects produced by the covering plates}
\label{sec:MeanFlowAndEJ-2}
In order to study the impact of the downwash motion over the disc surface, we recall that, as noted in \S\ref{sec:integralDragmeasurements}, the covering plates bear similarity to the streamwise-elongated rectangular roughness elements studied, for example, by \citet{hwang2018secondary}, \citet{vanderwel2019instantaneous} and \citet{von2021parametric}. The height of the covering plates is approximately $0.08\delta_{99}^*$, as given in table~\ref{tab:discScaling}. This relative height is comparable with the heights of roughness elements described in the literature. These investigations report that secondary streamwise vortices are induced near the top corners of the roughness elements. Elements with dimensions and spacings similar to our covering plates generate a secondary flow with a $-\avertime{v^*} = 0.02-0.03 U^*_{\infty}$ downwash motion on the valley side, a flow impingement on the vertical face of the element, and an upwash of comparable magnitude on the upper side of the element \citep{hwang2018secondary, vanderwel2015effects}. The mean-flow topology near the edges of our covering plates, reported in figure~\ref{fig:meanFlowContours}, is similar to the secondary-flow pattern produced by the roughness elements, with a comparable downwash and wall-normal penetration depth of $y/\delta_{99} = 0.8$. Unlike the canonical flows over elongated rectangular elements, however, our flow is modulated along the streamwise direction because of the disc motion.

Despite the streamwise modulation by the disc, the ridges introduce an intense spanwise inhomogeneity, which is the main cause of the formation of this Prandtl secondary flow of the second kind. This flow arises from spatial gradients in the Reynolds stress tensor and emerges even for slight spanwise heterogeneities \citep{bradshaw1987turbulent,anderson2015numerical,vanderwel2015effects}. Owing to the similarity between the two flows, it is plausible to expect an upwash motion of comparable magnitude on the top side of the cover. Similar secondary flows of comparable magnitude also form for rectangular ridges in internal channel flows, as reported by the DNS and experimental study of \citet{frohnapfel2024flow}. The downwash motion would therefore also occur if the test surface with the covering plates were utilised in the confined channel-flow configuration of \citet{Olivucci2021reduction}.

For roughness elements as wide as our covering plates, \citet{hwang2018secondary} and \citet{wangsawijaya2020effect} report that the spanwise width of the secondary flow is confined between $\pm (0.5-0.7)\delta^*_{99}$. The top view in figure~\ref{fig:meanFlowContours}(f) shows a similar pattern, i.e. the intense downwash motion only occurs over a small spanwise extent near the covering plates, $-0.1<z_d<0$ ($-0.5<z_d/\delta_{99}<0$). This result further supports the conjecture that the downwash motion is induced primarily by the secondary flow due to the covering plates. This secondary flow is also modulated by the disc flow there, which causes the streamwise inhomogeneous response of $\avertime{u}$ and $\avertime{v}$. In an outer-disc region towards the disc tip, where $-0.5<z_d<-0.1$, the less intense downwash motion, i.e. $\avertime{v}^+= -0.2$ shown in figure~\ref{fig:meanFlowContours}(f), is distributed across the disc surface. The influence of the covering plates is minimal in this region and the downwash motion is likely to be caused mainly by the disc rotation.

%====================================================================
\subsection{Statistics of velocity fluctuations over a rotating disc}
\label{sec:hoStatisticsTopology}

\begin{figure}
    \centering
    \includegraphics{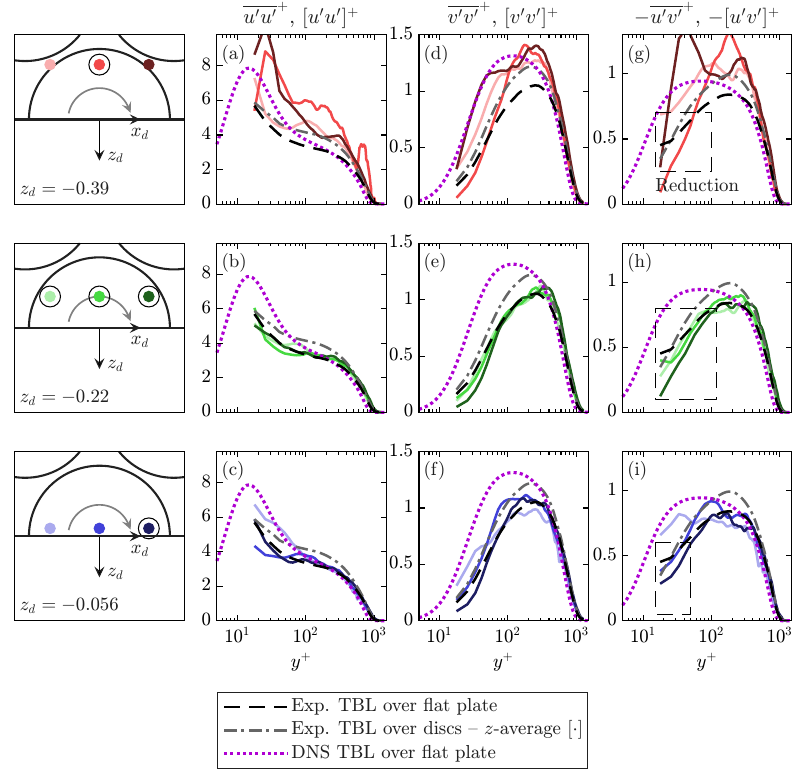}
    \caption{Wall-normal profiles of turbulent stresses, viscous scaled with $u_{\tau,d}^*$ of the flat-plate, of (a-c) streamwise normal stresses, (d-f) turbulent wall-normal normal stresses, and (g-i) Reynolds shear stresses. The dashed boxes in (g-i) show the near-wall region where the Reynolds shear stresses are reduced. Their $(x_d,z_d)$ locations are marked by the circles in the schematics in the left column. The layout and the line styles correspond to figure~\ref{fig:meanStatisticsProfiles}.}
    \label{fig:hoStatisticsProfiles}
\end{figure}

In this section, the influence of the test surface and of the disc motion on the second-order turbulence statistics is presented. The wall-normal profiles of $\avertime{u'u'}^+$, $\avertime{v'v'}^+$ and $-\avertime{u'v'}^+$ at different locations are shown in figure~\ref{fig:hoStatisticsProfiles}, for the TBLs over the flat plate and over the rotating-disc surface. DNS data by \citet{schlatter2010assessment} are also shown at a similar Reynolds number $Re_\tau = 830$. For $y^+ < 200$, all three measured quantities for the flat-plate case have a lower magnitude than those computed via DNS. This discrepancy is expected because of the spatial-averaging effect of the PIV, discussed in \S\ref{sec:EJPIV}. The effects of the disc motion can nevertheless be evaluated by comparing the coloured solid lines for the rotating-disc case to the black dashed lines for the flat-plate case.

%===============================================================================
\subsubsection{Modification of streamwise and wall-normal velocity fluctuations}

\begin{figure}
    \centering
    \includegraphics{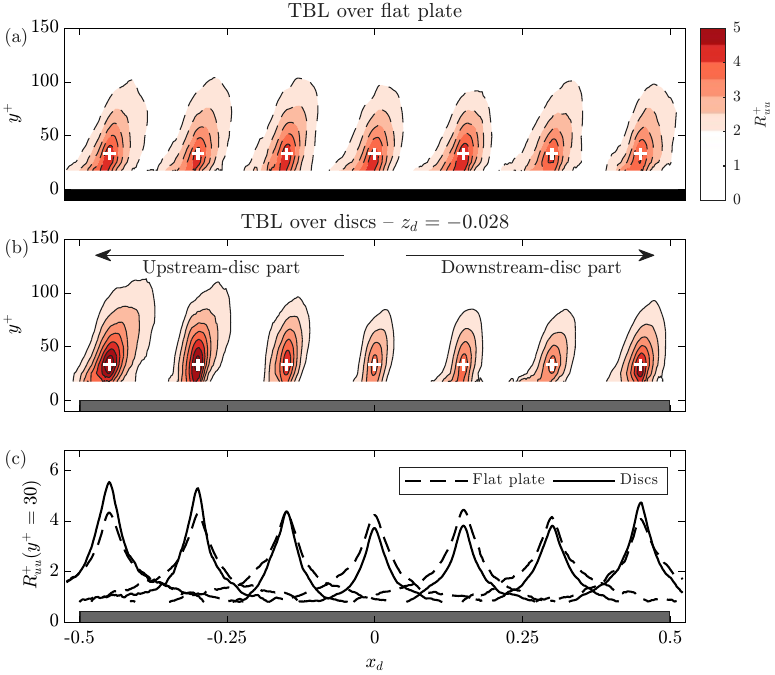}
    \caption{Spatial distribution of the two-point variance $R_{uu}^+$, defined in \eqref{eq:Ruu}. White plus markers indicate reference locations $x_{d,ref} = [-0.45:0.15:0.45]$ and $y_{ref}^+ = 30$. (a) Contour for the flat-plate case. (b) Contour for the rotating-disc case at $z_d=-0.028$. (c) Streamwise profiles at $y^+=30$ for the flat-plate (dashed line) and rotating disc (solid line) cases. Black and grey rectangles denote the flat-plate and the disc surfaces.}
    \label{fig:uu-covariances}
\end{figure}

Figure~\ref{fig:hoStatisticsProfiles}(c) shows that, at $z_d = -0.056$, i.e. at locations close to the covering plates, the streamwise-velocity variance $\avertime{u'u'}^+$ is enhanced on the upstream side of the disc for $y^+ < 200$. The profile at $(x_d,z_d) = (0.35,-0.056)$ is instead in close agreement with the flat-plate profile, while the profile at $(x_d,z_d) = (0,-0.056)$ shows a reduction. Figure~\ref{fig:hoStatisticsProfiles}(b) reports that, at $(x_d,z_d)= (0,-0.22)$, a reduction of $\avertime{u'u'}^+$ occurs for $y^+ < 200$. Figure~\ref{fig:hoStatisticsProfiles}(a) shows that, near the disc tips and between discs,  an enhancement of $\avertime{u'u'}^+$ occurs at any wall-normal location. To evaluate the spatial coherence of the turbulent fluctuations, the two-point covariance is computed,
\begin{equation}
    \label{eq:Ruu}
    R_{uu}(x,y) = \avertime{u'(x,y)u'(x_{ref},y_{ref})}.
\end{equation}
The covariance is computed at $y_{ref}^+ = 30$ at seven streamwise locations and is scaled in viscous units to quantify both the change in the covariance intensity and its spatial organisation. Figure~\ref{fig:uu-covariances} shows $R_{uu}^+$ at $z_d = -0.028$, which quantifies the intensity of the near-wall turbulent streaks. Figures~\ref{fig:uu-covariances}(a,b) compare the data for flat-plate TBL with the TBL over the rotating disc, while figure~\ref{fig:uu-covariances}(c) quantifies the streamwise coherence via $R^+_{uu}(x_d,y_{ref}^+ = 30)$ lines. Consistent with figure~\ref{fig:hoStatisticsProfiles}(c), when the disc is in motion, the variance is intensified in the upstream part (i.e. $x_d<0$), while the largest reduction occurs in the central region of the disc. When $x_d> -0.25$, the disc motion also renders the structures more compact along both the streamwise and wall-normal directions.

The wall-normal-velocity variance $\avertime{v'v'}^+$ shows slight deviations from the TBL over the flat plate across most of the disc surface, as shown in figures~\ref{fig:hoStatisticsProfiles}(e,f). Figure~\ref{fig:hoStatisticsProfiles}(d) reveals that, towards the disc tips, the variance $\avertime{v'v'}^+$ is strongly enhanced. This result is consistent with the localised mean outflow $\avertime{v}>0$ from the clearance gaps, as shown in figure~\ref{fig:meanFlowContours}(f). The strong enhancement of $\avertime{v'v'}^+$ near the disc edges (light and dark lines in figure~\ref{fig:hoStatisticsProfiles}d) may also indicate an unsteady transpiration through the gaps that would enhance the wall-normal velocity variance.
As the profiles of the streamwise and wall-normal velocity fluctuations at  $z_d = -0.39$, displayed in figures~\ref{fig:hoStatisticsProfiles}(a,d), both increase by more than 100\% at some wall-normal locations, the spanwise-averaged variances, indicated by the grey dash-dotted lines, exceed the flat-plate reference values.

%==========================================================================
\subsubsection{Modification of Reynolds shear stresses and drag-reduction effect}
\label{sec:uv}
The primary drag-reduction mechanism advanced in the DNS studies of wall turbulence over rotating discs is an overall reduction of the Reynolds stresses $-\avertimezed{u'v'}$ \citep{Ricco2013disks, Wise2014Oscilating, Olivucci2021reduction}. The terminology Reynolds stresses is hereafter adopted for the Reynolds shear stresses. While these numerical studies demonstrate a local increase in Reynolds stresses $-\avertime{u'v'}$ near the disc tips due to the interaction of wall turbulence with the high-shear flow between adjacent discs, the enhancement is offset by an intense attenuation of the stresses over the disc surface. The disc motion is therefore beneficial and an overall drag-reduction effect occurs.

Analogous to these DNS studies, figure~\ref{fig:hoStatisticsProfiles}(g) shows that $-\avertime{u'v'}^+$ increases near the disc tips, where the shear between rotating discs is intense. This enhancement is more significant than in the DNS studies, confirming the roughness effect caused by the clearance gaps, as discussed in \S\ref{sec:MeanFlowAndEJ-1}. Figures~\ref{fig:hoStatisticsProfiles}(g-i), however, also show that a significant reduction in $-\avertime{u'v'}^+$ occurs near the rotating-disc surface at various locations. These locations are indicated by the circled dots in the schematics in the left column of figure~\ref{fig:hoStatisticsProfiles}. The Reynolds stresses in figure~\ref{fig:hoStatisticsProfiles}(i) are attenuated near the covering plate at $(x_d,z_d) = (0.35,-0.056)$, where the disc moves towards the covering plate. The profiles in figure~\ref{fig:hoStatisticsProfiles}(h) at $z_d = -0.22$, i.e. centrally on the disc, show an attenuation up to $-\Delta \avertime{u'v'}^+=-0.4$ that extends to $y^+ = 200$ at all three streamwise locations. The Reynolds stresses in figure~\ref{fig:hoStatisticsProfiles}(g) show a reduction for $y^+<50$ only at $(x_d,z_d) = (0,-0.39)$, i.e. in a central position on the disc surface. These results demonstrate that, notwithstanding the $-\avertimezed{u'v'}$ enhancement due to the roughness effects, the rotation also causes a near-wall reduction of Reynolds stresses on the central part of the disc, providing the first experimental proof of the drag-reduction mechanism discussed in the DNS study of \citet{Olivucci2021reduction}.
\begin{figure}
    \centering
    \includegraphics{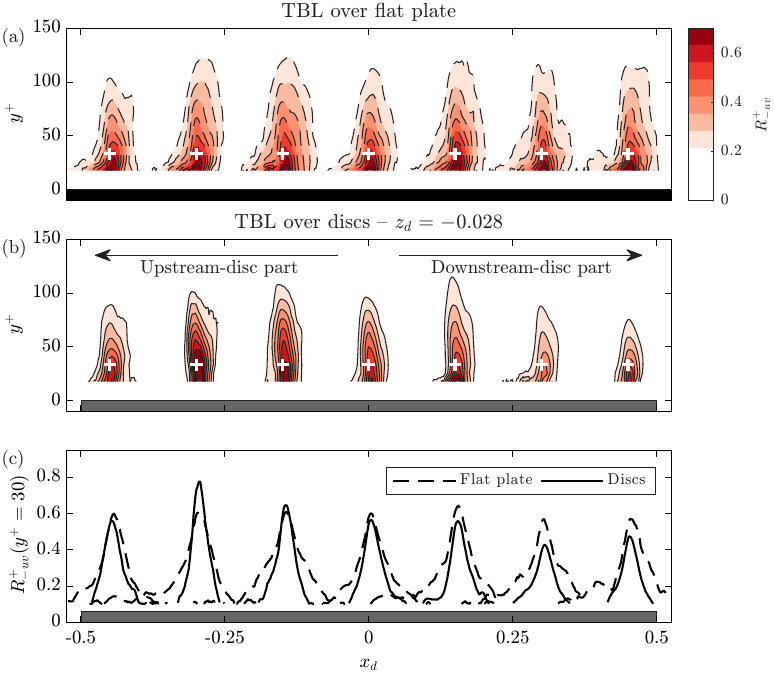}
    \caption{Spatial distribution of the two-point covariance $R_{-uv}^+$, defined in \eqref{eq:Ruv}. White plus markers indicate reference locations $x_{d,ref} = [-0.45:0.15:0.45]$ and $y_{ref}^+ = 30$. (a) Contour for the flat-plate case. (b) Contour for the rotating-disc case at $z_d=-0.028$. (c) Streamwise profiles at $y^+=30$ for the flat-plate (dashed line) and rotating disc (solid line) cases. Black and grey rectangles denote the flat-plate and the disc surfaces.}
    \label{fig:uv-covariances}
\end{figure}

Due to the three-dimensional flow caused by the discs, modifications of the Reynolds stress tensor and the coherent flow structures are expected \citep{moin1990direct,lozano2018modeling}. Figure~\ref{fig:uv-covariances} shows contours and profiles of the $R_{-uv}^+$ covariance, computed as
\begin{equation}
    \label{eq:Ruv}
    R_{-uv}(x,y) = -\avertime{u'(x,y)v'(x_{ref},y_{ref})},
\end{equation}
which represents the drag-producing structures, in the proximity of the covering plates at $z_d = -0.028$. When the disc is in motion, the covariance magnitude is intensified by up to 50\% in the upstream part, remains unvaried near the centre of the disc and is attenuated in the downstream part. The spatial coherence is intensely reduced by the disc motion, by up to half relative to the flat-plate case in both the $x$ and $y$ directions. This result provides strong evidence in support of the mechanism that the disc motion dampens the drag-producing turbulent structures.
\begin{figure}
    \centering
    \includegraphics{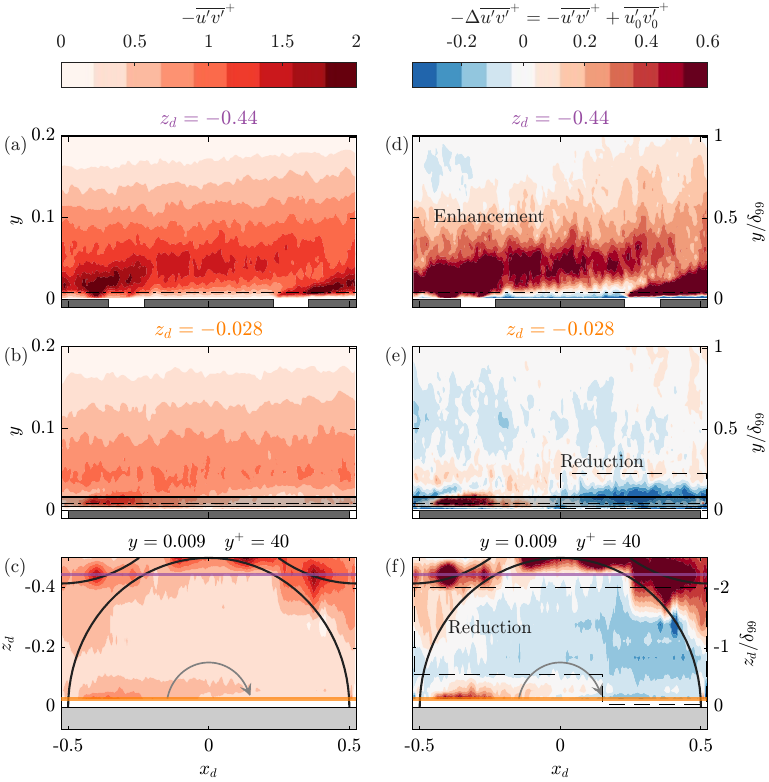}
    \caption{Contours of the (a,b,c) Reynolds stresses, $-\avertime{u'v'}^+$ and (d,e,f) their difference to the flat-plate reference, $-\Delta\avertime{u'v'}^+$. The flat-plate component is streamwise averaged to reduce the impact of statistical noise. The layout is the same as that of figure~\ref{fig:meanFlowContours}.}
    \label{fig:uvFlowContours}
\end{figure}

Figures~\ref{fig:uvFlowContours}(a-c) show the contours of the Reynolds stresses and figures~\ref{fig:uvFlowContours}(d-f) show the contours of the difference between these stresses in the rotating-disc case and the stresses in the flat-plate case. In the $x_d - y$ plane near the disc tips, figures~\ref{fig:uvFlowContours}(a,d) confirm that the surface-roughness effect and the shear between adjacent discs contribute to the enhancement of $-\avertime{u'v'}^+$ in the bulk of the flow. In agreement with the profile at $(x_d,z_d) = (0,-0.39)$, shown in figure~\ref{fig:hoStatisticsProfiles}(g), the Reynolds stresses are, however, reduced in the central part of the disc and close to the surface, i.e. for $y^+ < 35$.

The streamwise modulation of the Reynolds stresses due to the disc motion is clear in the plane at $z_d = -0.028$, shown in figures~\ref{fig:uvFlowContours}(b,e). On the downstream half of the disc, i.e. $x_d>0$, where its motion is towards the covering plate, the Reynolds stresses are attenuated by a maximum $-\Delta \avertime{u'v'}^+=-0.3$ up to $y^+ = 150$, confirming the result of figure~\ref{fig:uv-covariances}. The modification of the Reynolds stresses is in good agreement with that reported by \citet{Olivucci2021reduction}, who also show a reduction of $-\Delta\avertime{u'v'}^+ = -0.2$ distributed over the half-disc, with a wall-normal penetration of $y^+ = 80$.

Figures~\ref{fig:uvFlowContours}(c,f) show the Reynolds stresses and their attenuation in an $x_d - z_d$ plane at $y^+ = 40$ (this wall-normal location is indicated by the dash-dotted line in panels (a,b,d,e)). The region where the stresses reduce extends over nearly the entire half-disc surface, as marked by the enclosed dashed line in figure~\ref{fig:uvFlowContours}(f). The results of figure~\ref{fig:uvFlowContours}(f) further confirm that the disc motion induces the indirect drag-reduction mechanism based on the Reynolds-stress reduction, as put forward by \citet{Olivucci2021reduction}. The top view of $-\Delta\avertime{u'v'}^+$ at $y^+ = 40$ reveals that, near the covering plate and in the upstream part of the disc, a narrow spanwise region of $-\Delta \avertime{u'v'}^+>0$ occurs. This enhancement is qualitatively similar to the spatial distribution of skin-friction drag reduction reported in \citet{Olivucci2021reduction} in their figure 13c. Analogous to our experimental results, \citet{Olivucci2021reduction} found a thin, streamwise-stretched region of drag increase in the upstream part, i.e. $x_d<0$, where the disc moves away from the covering geometry. 

%===============================================================================
\subsection{Elnahhas-Johnson skin-friction budget for flow over a rotating disc}
\label{sec:CfBudget}

\begin{figure}
    \centering
    \includegraphics{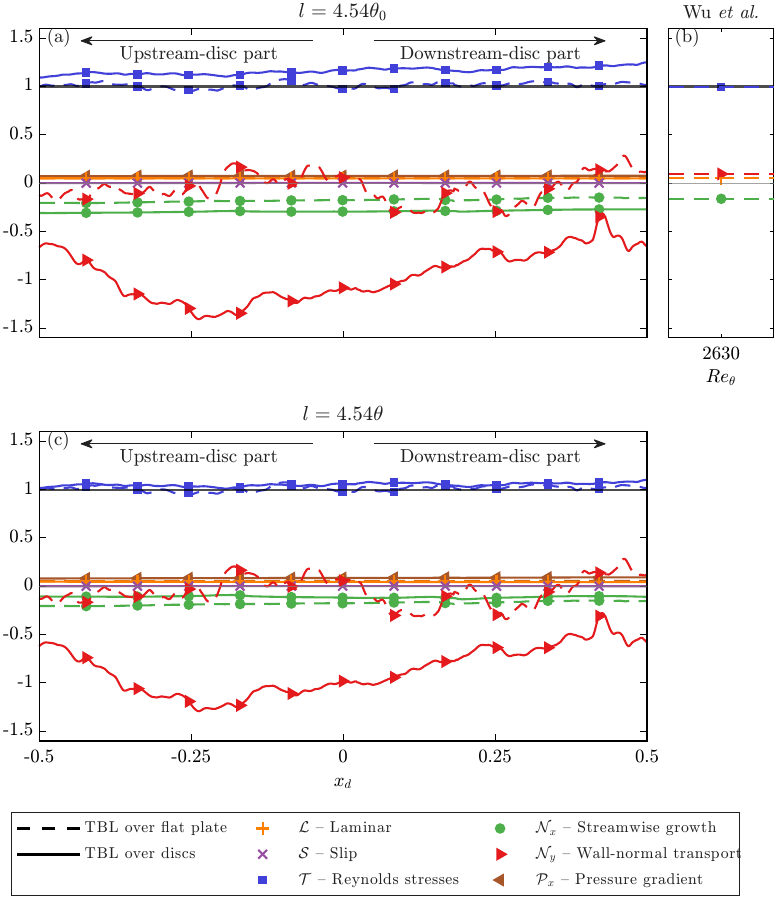}
    \caption{Skin-friction budget computed for the extended EJ identity \eqref{eq:EJ} using (a) $l(x)=4.54 \theta_0(x)$ and (c) $l(x)=4.54\theta(x)$. Lines for TBL over the flat plate are indicated by dashed lines and lines for TBL over the half-disc are indicated by solid lines.
    (b) Flat-plate TBL reference values at comparable $Re_\theta$ based on the data in \citet{wu2017transitional}. All terms are normalised by the $C_{f,d}/2$ of the flat-plate TBL (refer to table~\ref{tab:conditionsPIV}), as denoted by the horizontal black line at unity.}
    \label{fig:EJcomponents}
\end{figure}

The impact of the test surface and disc rotation on the terms in the EJ identity \eqref{eq:EJ} is discussed. Figure~\ref{fig:EJcomponents} presents the effect of the disc rotation on the EJ terms at $Re_{\tau,d} = 900$, normalised by $C_{f,d}/2$ of the flat-plate TBL. The momentum thickness Reynolds number is $Re_{\theta,d}= 2640-2850$ across the field of view, which gives $l(x) \approx 0.1$. Our choice of $l(x)$ affects the comparison between the flat-plate and rotating-disc cases. Using $l(x)=4.54\theta_0(x)$, i.e. the same flat-plate reference $\theta_0$ for both cases, highlights absolute changes with respect to the flat-plate TBL. Using the respective $\theta$ of each case, $l(x)=4.54\theta(x)$ accounts for the differences in TBL development by comparing to laminar boundary layers at equivalent $Re_\theta$ \citep{Elnahhas_Johnson_2022}. Figures~\ref{fig:EJcomponents}(a,c) show the EJ budget terms for these choices of $l(x)$. The dashed lines denote the flat-plate data, while the solid lines denote the rotating-disc data.

%==============================================================================
\subsubsection{Validation of skin-friction budget for flow over the flat plate}
The terms of the EJ identity computed using our experimental data agree very well with the terms computed by EJ, shown in their figures~9 and 10. They used numerical data for TBL from H-type and bypass transition simulations \citep{lozano2018modeling,zaki2013streaks,wu2017transitional} and from fully-turbulent recycling-rescaling simulations \citep{sillero2013one}. Figure~\ref{fig:EJcomponents}(b) shows the quantitative agreement between our results and those of EJ based on the numerical data of \citet{wu2017transitional} at a comparable Reynolds number $Re_\theta = 2630$. The deviations are due to the experimental uncertainty, as discussed in \S\ref{sec:EJPIV}. The term $\mathcal{N}_{y}$ displays the largest streamwise variation as expected because of the large uncertainty associated with the computation of $\avertimezed{v}$. Its streamwise-averaged value is nevertheless almost null, in agreement with $\mathcal{N}_{y}=0.1$ computed by EJ. The term $\mathcal{N}_{x} =-0.2$ is streamwise independent and matches the value computed by EJ very well. Our term $\mathcal{T}$ is close to unity, thus matching the values reported by EJ, i.e. $\mathcal{T}=1$ for the transition simulations \citep{lozano2018modeling,zaki2013streaks,wu2017transitional} and $\mathcal{T}=1.05$ for recycling-rescaling simulations \citep{sillero2013one}.
Figure \ref{fig:EJcomponents} also shows that the pressure-gradient term $\mathcal{P}_x$ is small, of the order of $\mathcal{L}$, because of the minimal variation of $U_\infty(x)$ along the streamwise direction.

%=================================================================
\subsubsection{Skin-friction budget for flow over a rotating disc}
The effect of the discs relative to the flat-plate reference, based on $l(x) = 4.54\theta_0(x)$, is shown in figure~\ref{fig:EJcomponents}(a). In the rotating-disc case, the direct-slip term $\mathcal{S} = -\mathcal{L}(U_w/U_\infty)$is negligible compared to the other terms because both its contributing terms are small, i.e. the slip velocity is $(U_w/U_\infty)=0.07$ and the laminar term is $\mathcal{L}=0.06$, as discussed in \S\ref{sec:slipDrag}. The effect of $\mathcal{P}_x$ remains small for the flow over the discs.

The non-parallel term $\mathcal{N}_y$ is significantly enhanced when the discs are in motion. The streamwise-velocity deficit $1-\avertimezed{u}$ and the negative wall-normal velocity $\avertimezed{v}$, engendered by the complexity of the test surface as discussed in \S\ref{sec:mean-flow-modification}, cause $\mathcal{N}_y$ to increase in absolute value, thus contributing to the drag-reduction effect reported by the purple symbols in figure~\ref{fig:DeltaCD}(b). The streamwise inhomogeneity of $\mathcal{N}_y$ is related to the large-scale modulation produced by the disc motion. In absolute value, it is large when the disc moves spanwise away from the covering plates, for $x_d<0$, and it reaches its smallest value along the streamwise direction when the disc moves spanwise towards the covering plates, for $x_d>0$. This significant streamwise flow dependence was not observed in the DNS simulations of \citet{Olivucci2021reduction}, where the covering plates were absent. 

\citet[in their figure 11]{hwang2018secondary} also reported a wall-shear stress that is lower than the overall averaged shear stress near the lower parts of their roughness elements, due to the secondary streamwise-elongated vortices forming near the edges of the elements. They also found a larger wall-shear stress than the overall average over the horizontal upper sides of the roughness elements. This increase offsets the reduction on the lower side, causing an overall drag increase.
Because of the similarities between these two flows, a friction enhancement is also likely to occur over the top part of our covering plates ($z_d>0$). This effect is, however, not quantified in the EJ budget because the covering plates are not part of the domain of integration. The overall balance would thus result in a drag increase, documented by the positive $\Delta C_{D,cover}$ in figure~\ref{fig:DeltaCD} and discussed in \S\ref{sec:integralDragmeasurements}.

The Reynolds-stress term $\mathcal{T}$ is enhanced in figure \ref{fig:EJcomponents}(a), implying that the overall change of the Reynolds stresses increases the drag with respect to the flat-plate case. This result contrasts with the findings from numerical investigations of actively and passively rotating discs, i.e. the decrease in $-\avertimezed{u'v'}$ is the primary mechanism of drag reduction \citep{Ricco2013disks,Wise2014Oscilating, olivucci2019turbulent,Olivucci2021reduction}. As amply discussed in \S\ref{sec:uv}, the increase of the Reynolds stresses arises from the test geometry, i.e. the covering plates, the disc housings and the radial gaps between the discs and their housings, all features that are absent in the idealised configuration considered in the DNS studies. Figure \ref{fig:EJcomponents}(a) further shows that the negative $\mathcal{N}_x$ is enhanced by an amount similar to that of $\mathcal{T}$ with respect to the flat-plate case.

Figure~\ref{fig:EJcomponents}(c) reveals how using $l(x)=4.54 \theta(x)$ instead adjusts the budget terms for dissimilarities in streamwise growth. The terms $\mathcal{T}$ and $\mathcal{N}_x$ in the rotating-disc case display a good agreement with the respective flat-plate data when scaled in this way. This result indicates that the enhancement of $\mathcal{T}$ due to the roughness effect in figure~\ref{fig:EJcomponents}(a) is largely cancelled by the increase in streamwise boundary-layer growth and is similar to the flat-plate $\mathcal{T}$ when differences in streamwise-development (i.e. $Re_\theta$) are accounted for. The qualitative trend in the other budget terms remains unvaried. These results highlight the importance of understanding the choice of the length scale $l(x)$ when interpreting EJ identity results in non-canonical cases.

The wall-transpiration term $\mathcal{V}_w$ in \eqref{eq:EJ} has a negligible impact on the balance because $\avertime{v}(y=0)$ is estimated to be less than 0.05 at $y=0.016$ (figure~\ref{fig:meanFlowContours}f) and the spanwise extent of the disc clearance is much smaller than the spanwise interval of integration. The velocity $\avertime{u}(y=0)$, estimated to be comparable to the disc tip velocity, does not impact the computation of $\mathcal{V}_w$ because it is much smaller than unity.

%================
\section{Summary}
\label{sec:conclusions}

In this study, we have extended the proof-of-concept of a single passively rotating disc developed by \citet{Koch2013Drag} by investigating the effect of thirty-two passively rotating discs on spatially developing turbulent boundary layers. The discs were flush to the wall surface and mounted on bearings located in internal housings. Spanwise halves of the discs were covered by thin plates. The discs were set in steady motion by the torque exerted by the surface-integrated wall-shear stress of the turbulent boundary layer flowing on the exposed parts of the discs.

We used direct force measurements, velocity measurements of the disc rotation and spatially resolved PIV velocity measurements to study how the disc motion contributes to the drag change over the test surface. The drag-modifying mechanisms were elucidated further by extending the skin-friction identity discovered by \citet{Elnahhas_Johnson_2022} to our passively-rotating disc case. The drag measurements, conducted in the range $Re_{\tau,avg} = 880 - 1460$, revealed that the disc motion causes a 3\% drag reduction with respect to the turbulent flow over the test surface with stationary discs. This result is the first experimental proof that rotating discs can reduce drag. The test surface, however, causes the overall drag to increase with respect to the flat-plate boundary layer. The drag increase is attributed to the secondary flow caused by the covering plates and to a roughness effect produced by the disc housings, specifically by the clearance gaps around the discs. The direct disc-slip contribution to the change in drag, computed using the Elnahhas-Johnson identity, is negligible, implying that the drag reduction is caused indirectly via a modification of the near-wall turbulence. 
The covariances of the velocity fluctuations and the Reynolds stresses are enhanced in the proximity of the disc tips because of the roughness effect and the shear between adjacent discs. 
The overall increment of the Reynolds stresses is responsible for the drag increase, as quantified by the related integral term in the Elnahhas-Johnson identity.
% The identity also shows that a drag decrease caused by an enhanced boundary-layer growth largely cancels the drag increase from the Reynolds stresses. 
Despite the overall drag increment by the Reynolds stresses, reductions in magnitude and spatial coherence of the Reynolds stresses were measured near the disc surface in the central part of the disc, proving that the disc motion alters the wall turbulence beneficially. 

The flow over the rotating-disc surface causes a strong mean-flow modulation along the streamwise and spanwise directions. These flow changes were not reported in previous numerical studies simulating idealised flow configurations because the covering plates and disc housings were absent. An intense downwash motion near the covering plates contributes to a large local drag-reducing effect via a term in the Elnahhas-Johnson identity that is related to the wall-normal transport of momentum deficit. Since the flow over the covering plates resembles the secondary flow over streamwise-aligned rectangular ridges, this downwash motion is expected to be counteracted by an upwash motion on top of the covering plates. 
% This upwash motion is responsible for the overall drag enhancement caused by the covering plates. 
Away from the covering plates and in the central part of the disc, where the influence of the secondary flow is negligible, a less intense downwash motion occurs in concurrence with the near-wall reduction of the Reynolds stresses, which is responsible for a distributed drag-reducing effect.

Our study emphasises the importance of considering the geometry required for implementing the passively rotating discs. Several open questions remain regarding the flow physics attributed to the test surface. These include the interaction between the near-surface disc flow and the secondary flow over the covering plates, and the drag-reduction effect produced by the reduction of Reynolds stresses. 
It would also be of interest to measure the streamwise and spanwise pressure gradients within the shear flow induced by the disc rotation on the disc surfaces and between discs. Future work should also involve measurements of the spanwise velocity for the accurate computation of terms $\mathcal{N}_z$ and $\mathcal{I}_z$ in the Elnahhas-Johnson identity in order to obtain the full skin-friction budget. Direct numerical simulations of the flow engendered by the complete test surface would be a useful tool to obtain full three-dimensional flow fields.

The drag penalties produced by the disc housings and the covering plates are of the same order as the 3\% drag reduction caused by the disc motion. This result is promising given that, for typical active control methods, the actual power requirement exceeds the beneficial power savings by several orders of magnitude. Future work should focus on understanding the mechanisms that cause the drag enhancement and reduction with a view to optimising the design and reducing the added drag given by the secondary flows and the roughness effects.

%=============================
\backsection[Acknowledgements]{We express our gratitude to Olaf W.\:G. van Campenhout for his support in realising the experimental model and his enduring enthusiasm during this investigation. We also express our gratitude to Michiel van Nesselrooij and Friso H. Hartog for their helpful discussions. The support from the technical staff at the Flow Physics \& Technology laboratory at the Delft University of Technology, namely, Stefan Bernardy, Emiel Langedijk, and Gert-Jan Berends, is gratefully acknowledged. M.K. is appreciative of the support from Dr Bas W. van Oudheusden and Dr Ferry F.\:J. Schrijer. We also thank Professor Daniel Chung and Professor Nicholas Hutchins at the University of Melbourne for their useful comments and suggestions that have improved the quality of our work.}
\backsection[Funding]{This research was financially supported by the Dutch Enterprise Agency (RvO) under grant number TSH21002. Part of P.R.'s research was financially supported by the Taiho Kogyo Tribology Research Foundation under grant number 191008.}
\backsection[Declaration of interests]{The authors report no conflict of interest.}
\backsection[Data availability statement]{A dataset of the results (figures~\ref{fig:DeltaCD} to \ref{fig:EJcomponents}) is made available \href{http://doi.org/10.4121/ae009f06-b3f4-43fd-8833-eacba80a7309}{online} \citep{Dataset}.}
\backsection[Author ORCIDs]{Max W. Knoop, https://orcid.org/0009-0008-8848-3006; P. Ricco, https://orcid.org/0000-0003-1537-1667}

\bibliographystyle{jfm}
\bibliography{jfm}

@article{Nesselrooij_development_2022,
   author = {van Nesselrooij, M. and van Campenhout, O.W.G. and van Oudheusden, B.W. and Schrijer, F.F.J. and Veldhuis, L.L.M.},
   title = {Development of an experimental apparatus for flat plate drag measurements and considerations for such measurements},
   journal = {Meas. Sci. Technol.},
   volume = {33},
   number = {5},
DOI = {10.1088/1361-6501/ac527f},
   year = {2022},
   type = {Journal Article}
}

@InProceedings{koch-kozulovic-2014,
  Author         = {Koch, H. and Kozulovic, D.},
  Title          = {Influence of Geometry Variations on the Boundary Layer Control with a Passively Moving Wall},
  BookTitle      = {AIAA SciTech Forum, 52nd Aero. Sc. Meeting},
  Pages          = {0401},
  year           = 2014
}

@article{westerweel2005universal,
  title={Universal outlier detection for {PIV} data},
  author={Westerweel, J. and Scarano, F.},
  journal={Exp. Fluids.},
  volume={39},
  pages={1096--1100},
  year={2005},
  publisher={Springer},
DOI = {10.1007/s00348-005-0016-6}
}

@inproceedings{keefe1997normal,
  title={A normal vorticity actuator for near-wall modification of turbulent shear flows},
  author={Keefe, L.},
  booktitle={35th Aerospace Sciences Meeting and Exhibit},
  pages={547},
  year={1997}
}

@article{Ricco2013disks,
   author = {Ricco, P. and Hahn, S.},
   title = {Turbulent drag reduction through rotating discs},
   journal = {J. Fluid Mech.},
   volume = {722},
   pages = {267-290},
   ISSN = {0022-1120},
   year = {2013},
   type = {Journal Article},
DOI = {10.1017/jfm.2013.92}
}

@inproceedings{Koch2013Drag,
   author = {Koch, H. and Kozulovic, D.},
   title = {Drag reduction by boundary layer control with passively moving wall},
   booktitle = {Fluids Engineering Division Summer Meeting},
   publisher = {American Society of Mechanical Engineers},
   volume = {55553},
   pages = {V01BT15A004},
   ISBN = {0791855554},
   year = {2013},
   type = {Conference Proceedings}
}

@article{Olivucci2021reduction,
   author = {Olivucci, P. and Wise, D.J. and Ricco, P.},
   title = {Reduction of turbulent skin-friction drag by passively rotating discs},
   journal = {J. Fluid Mech.},
   volume = {923},
   ISSN = {0022-1120},
   year = {2021},
   type = {Journal Article},
DOI = {10.1017/jfm.2021.533}
}

@article{Wise2014SpinningConfig,
   author = {Wise, D.J. and Alvarenga, C. and Ricco, P.},
   title = {Spinning out of control: Wall turbulence over rotating discs},
   journal = {Phys. Fluids},
   volume = {26},
   number = {12},
   pages = {125107},
   ISSN = {1070-6631},
   year = {2014},
   type = {Journal Article},
DOI = {10.1063/1.4903973}
}

@article{Wise2014Oscilating,
   author = {Wise, D.J. and Ricco, P.},
   title = {Turbulent drag reduction through oscillating discs},
   journal = {J. Fluid Mech.},
   volume = {746},
   pages = {536-564},
   ISSN = {0022-1120},
   year = {2014},
   type = {Journal Article},
DOI = {10.1017/jfm.2014.122}
}

@article{Elnahhas_Johnson_2022, 
title={On the enhancement of boundary layer skin friction by turbulence: an angular momentum approach}, 
author={Elnahhas, A. and Johnson, P.L.},
volume={940}, 
DOI={10.1017/jfm.2022.264}, 
journal={J. Fluid Mech.}, 
year={2022}, 
pages={A36}
}

@article{karman1921laminare,
  title={{\"U}ber laminare und turbulente Reibung},
  author={von K{\'a}rm{\'a}n, T},
  journal={Zeitschrift f{\"u}r Angewandte Mathematik und Mechanik},
  volume={1},
  number={4},
  pages={233--252},
  year={1921},
  publisher={Wiley Online Library}
}

@article{vanderwel2019instantaneous,
  title={The instantaneous structure of secondary flows in turbulent boundary layers},
  author={Vanderwel, C. and Stroh, A. and Kriegseis, J. and Frohnapfel, B. and Ganapathisubramani, B.},
  journal={J. Fluid Mech.},
  volume={862},
  pages={845--870},
  year={2019},
  publisher={Cambridge University Press},
DOI = {10.1017/jfm.2018.955}
}

@article{hwang2018secondary,
  title={Secondary flows in turbulent boundary layers over longitudinal surface roughness},
  author={Hwang, H.G. and Lee, J.H.},
  journal={Phys. Rev. Fluids},
  volume={3},
  number={1},
  pages={014608},
  year={2018},
  publisher={APS},
DOI = {10.1103/PhysRevFluids.3.014608}
}

@article{von2021parametric,
  title={Parametric study on ridges inducing secondary motions in turbulent channel flow},
  author={von Deyn, L.H. and Gatti, D. and Frohnapfel, B. and Stroh, A.},
  journal={Proc. Appl. Math. Mech.},
  volume={20},
  number={1},
  pages={e202000139},
  year={2021},
  publisher={Wiley Online Library},
DOI = {10.1002/pamm.202000139}
}

@article{chauhan2009criteria,
  title={Criteria for assessing experiments in zero pressure gradient boundary layers},
  author={Chauhan, K.A. and Monkewitz, P.A. and Nagib, H.M.},
  journal={Fluid Dyn. Res.},
  volume={41},
  number={2},
  pages={021404},
  year={2009},
  publisher={IOP Publishing},
DOI = {10.1088/0169-5983/41/2/021404}
}

@Article{bechert-etal-1996,
  Author         = {Bechert, D.W. and Hage, W. and Brusek, M.},
  Title          = {Drag reduction with the slip wall},
  Journal        = {AIAA J.},
  Volume         = {34 (5)},
  Pages          = {1072},
  catalog        = {1311},
  year           = 1996
}

@Article{quadrio-ricco-2004,
  Author         = {Quadrio, M. and Ricco, P.},
  Title          = {Critical assessment of turbulent drag reduction
                   through spanwise wall oscillations},
  Journal        = {J. Fluid Mech.},
  Volume         = {521},
  Pages          = {251--271},
  year           = 2004
}

@article{chung2021predicting,
  title={Predicting the drag of rough surfaces},
  author={Chung, D. and Hutchins, N. and Schultz, M.P. and Flack, K.A.},
  journal={Annu. Rev. Fluid Mech.},
DOI = {10.1146/annurev-fluid-062520-115127},
  volume={53},
  number={1},
  pages={439--471},
  year={2021},
  publisher={Annual Reviews}
}

@article{hartog2024turbulent,
  title={Turbulent boundary layers over substrates with streamwise-preferential permeability},
  author={Hartog, F.H. and van Nesselrooij, M. and van Campenhout, O.W.G. and Schrijer, F.F.J. and van Oudheusden, B.W. and Masania, K.},
  journal={Phys. Rev. Fluids},
DOI = {10.1103/PhysRevFluids.9.114602},
  volume={9},
  number={11},
  pages={114602},
  year={2024},
  publisher={APS}
}

@article{van2023experimental,
  title={Experimental and numerical investigation into the drag performance of dimpled surfaces in a turbulent boundary layer},
  author={van Campenhout, O.W.G. and van Nesselrooij, M. and Lin, Y.Y. and Casacuberta, J. and van Oudheusden, B.W. and Hickel, S.},
  journal={Int. J. Heat Fluid Flow},
DOI = {10.1016/j.ijheatfluidflow.2023.109110},
  volume={100},
  pages={109110},
  year={2023},
  publisher={Elsevier}
}

@article{scarano2000advances,
  title={Advances in iterative multigrid {PIV} image processing},
  author={Scarano, F. and Riethmuller, M.L.},
  journal={Exp. Fluids},
  volume={29},
DOI = {10.1007/s003480070007},
  number={Suppl 1},
  pages={S051--S060},
  year={2000},
  publisher={Springer}
}

@article{olivucci2019turbulent,
  title={Turbulent drag reduction by rotating rings and wall-distributed actuation},
  author={Olivucci, P. and Ricco, P. and Aghdam, S.K.},
  journal={Phys. Rev. Fluids},
  volume={4},
  number={9},
  pages={093904},
  year={2019},
  publisher={APS},
DOI = {10.1103/PhysRevFluids.4.093904}
}

@article{fukagata2002contribution,
  title={Contribution of {R}eynolds stress distribution to the skin friction in wall-bounded flows},
  author={Fukagata, K. and Iwamoto, K. and Kasagi, N.},
  journal={Phys. Fluids},
  volume={14},
  number={11},
  pages={L73--L76},
  year={2002},
  publisher={American Institute of Physics},
DOI = {10.1063/1.1516779}
}

@article{lee2016validating,
  title={Validating under-resolved turbulence intensities for {PIV} experiments in canonical wall-bounded turbulence},
  author={Lee, J.H. and Kevin and Monty, J.P. and Hutchins, N.},
  journal={Exp. Fluids},
  volume={57},
  pages={1--11},
  year={2016},
  publisher={Springer},
DOI = {10.1007/s00348-016-2209-6}
}

@article{adrian1997dynamic,
  title={Dynamic ranges of velocity and spatial resolution of particle image velocimetry},
  author={Adrian, R.J.},
  journal={Meas. Sci. Technol.},
  volume={8},
  number={12},
  pages={1393},
  year={1997},
  publisher={IOP Publishing},
DOI = {10.1088/0957-0233/8/12/003}
}

@article{wilson2013uncertainty,
  title={Uncertainty on {PIV} mean and fluctuating velocity due to bias and random errors},
  author={Wilson, B.M. and Smith, B.L.},
  journal={Meas. Sci. Technol.},
  volume={24},
  number={3},
  pages={035302},
  year={2013},
  publisher={IOP Publishing},
DOI = {10.1088/0957-0233/24/3/035302}
}

@article{abbassi2017skin,
  title={Skin-friction drag reduction in a high-{R}eynolds-number turbulent boundary layer via real-time control of large-scale structures},
  author={Abbassi, M.R. and Baars, W.J. and Hutchins, N. and Marusic, I.},
  journal={Int. J. Heat Fluid Flow},
  volume={67},
  pages={30--41},
  year={2017},
  publisher={Elsevier},
DOI = {10.1016/j.ijheatfluidflow.2017.05.003}
}

@article{dacome2024opposition,
  title={Opposition flow control for reducing skin-friction drag of a turbulent boundary layer},
  author={Dacome, G. and M{\"o}rsch, R. and Kotsonis, M. and Baars, W.J.},
  journal={Phys. Rev. Fluids},
  volume={9},
  number={6},
  pages={064602},
  year={2024},
  publisher={APS},
DOI = {10.1103/PhysRevFluids.9.064602}
}

@article{vanNesselrooij2016drag,
  title={Drag reduction by means of dimpled surfaces in turbulent boundary layers},
  author={van Nesselrooij, M. and Veldhuis, L.L.M. and van Oudheusden, B.W. and Schrijer, F.F.J},
  journal={Exp. Fluids},
  volume={57},
  pages={1--14},
  year={2016},
  publisher={Springer}, 
DOI = {10.1007/s00348-016-2230-9}
}

@article{quadrio_streamwise-travelling_2009,
	title = {Streamwise-travelling waves of spanwise wall velocity for turbulent drag reduction},
	volume = {627},
	issn = {1469-7645},
        doi = {10.1017/S0022112009006077},
	pages = {161--178},
	journal = {J. Fluid Mech.},
	author = {Quadrio, M. and Ricco, P. and Viotti, C.},
	year = {2009},
}

@article{quadrio_laminar_2011,
	title = {The laminar generalized {S}tokes layer and turbulent drag reduction},
	volume = {667},
	issn = {0022-1120},
	doi = {10.1017/s0022112010004398},
	pages = {135--157},
	journal = {J. Fluid Mech.},
	author = {Quadrio, M. and Ricco, P.},
	year = {2011},
}

@article{Touber_near-wall_2012,
   author = {Touber, E. and Leschziner, M.A.},
   title = {Near-wall streak modification by spanwise oscillatory wall motion and drag-reduction mechanisms},
   journal = {J. Fluid Mech.},
   volume = {693},
   pages = {150-200},
   ISSN = {0022-1120},
   DOI = {10.1017/jfm.2011.507},
   year = {2012},
   type = {Journal Article}
}

@article{agostini_turbulence_2015,
  title={The turbulence vorticity as a window to the physics of friction-drag reduction by oscillatory wall motion},
  author={Agostini, L. and Touber, E. and Leschziner, M.A.},
  journal={Int. J. Heat Fluid Flow},
  volume={51},
  pages={3--15},
  year={2015},
    DOI = {10.1016/j.ijheatfluidflow.2014.08.002},
  publisher={Elsevier}
}

@article{auteri2010experimental,
  title={Experimental assessment of drag reduction by traveling waves in a turbulent pipe flow},
  author={Auteri, F. and Baron, A. and Belan, M. and Campanardi, G. and Quadrio, M.},
  journal={Phys. Fluids},
  volume={22},
  number={11},
  year={2010},
  publisher={AIP Publishing},
DOI = {10.1063/1.3491203}
}

@phdthesis{koch2014reduzierung,
    author={Koch, H.J.},
    title={Reduzierung des Str{\"o}mungswiderstandes durch partielle Gleitoberfl{\"a}chen},
    school = {Technical University of Braunschweig},
    year = {2014}
}

@Article{jung-mangiavacchi-akhavan-1992,
  Author         = {Jung, W.J. and Mangiavacchi, N. and Akhavan, R.},
  Title          = {Suppression of turbulence in wall-bounded flows by
                   high-frequency spanwise oscillations},
  Journal        = {Phys. Fluids A},
  Volume         = {4},
  Number         = {8},
  Pages          = {1605-1607},
  catalog        = {403},
  year           = 1992
}

@article{carrasco2024experimental,
  title={Experimental Investigation into the Drag Performance of Chevron-Shaped Protrusions in Wall-Bounded Turbulence},
  author={Carrasco Grau, J. and van Campenhout, O.W.G. and Hartog, F.H. and van Nesselrooij, M. and Baars, W.J. and Schrijer, F.F.J.},
  journal={Flow Turbul. Combust.},
  volume={113},
  number={1},
  pages={159--175},
  year={2024},
  publisher={Springer}, 
DOI = {10.1007/s10494-023-00451-0}
}

@article{hervet2003flow,
  title={Flow with slip at the wall: from simple to complex fluids},
  author={Hervet, H. and L{\'e}ger, L.},
  journal={Comptes rendus. Physique},
  volume={4},
  number={2},
  pages={241--249},
  year={2003},
doi = {10.1016/S1631-0705(03)00047-1}
}

@article{rastegari2019drag,
  title={On drag reduction scaling and sustainability bounds of superhydrophobic surfaces in high {R}eynolds number turbulent flows},
  author={Rastegari, A. and Akhavan, R.},
  journal={J. Fluid Mech.},
  volume={864},
  pages={327--347},
  year={2019},
  publisher={Cambridge University Press},
doi = {10.1017/jfm.2018.1027}
}

@article{medjnoun2020effects,
  title={Effects of heterogeneous surface geometry on secondary flows in turbulent boundary layers},
  author={Medjnoun, T. and Vanderwel, C. and Ganapathisubramani, B.},
  journal={J. Fluid Mech.},
  volume={886},
  pages={A31},
  year={2020},
  publisher={Cambridge University Press},
DOI = {10.1017/jfm.2019.1014}
}

@article{min2004effects,
  title={Effects of hydrophobic surface on skin-friction drag},
  author={Min, T. and Kim, J.},
  journal={Phys. Fluids},
  volume={16},
  number={7},
  pages={L55--L58},
  year={2004},
  publisher={American Institute of Physics}, 
DOI = {10.1063/1.1755723}
}

@article{schlatter2010assessment,
  title={Assessment of direct numerical simulation data of turbulent boundary layers},
  author={Schlatter, P. and {\"O}rl{\"u}, R.},
  journal={J. Fluid Mech.},
  volume={659},
  pages={116--126},
  year={2010},
  publisher={Cambridge University Press},
DOI = {10.1017/S0022112010003113}
}

@article{bird2018experimental,
  title={Experimental control of turbulent boundary layers with in-plane travelling waves},
  author={Bird, J. and Santer, M. and Morrison, J.F.},
  journal={Flow Turbul. Combust.},
  volume={100},
  pages={1015--1035},
  year={2018},
  publisher={Springer},
DOI = {10.1007/s10494-018-9926-2}
}

@article{gatti2015experimental,
  title={Experimental assessment of spanwise-oscillating dielectric electroactive surfaces for turbulent drag reduction in an air channel flow},
  author={Gatti, D. and G{\"u}ttler, A. and Frohnapfel, B. and Tropea, C.},
  journal={Exp. Fluids},
  volume={56},
  pages={1--15},
  year={2015},
  publisher={Springer},
DOI = {10.1007/s00348-015-1983-x}
}

@article{wu2017transitional,
  title={Transitional--turbulent spots and turbulent--turbulent spots in boundary layers},
  author={Wu, X. and Moin, P. and Wallace, J.M. and Skarda, J. and Lozano-Dur{\'a}n, A. and Hickey, J.-P.},
  journal={Proc. Natl. Acad. Sci. USA},
  volume={114},
  number={27},
  pages={E5292--E5299},
  year={2017},
  publisher={National Academy of Sciences},
DOI = {10.1073/pnas.1704671114}
}

@misc{Dataset,
doi = {10.4121/ae009f06-b3f4-43fd-8833-eacba80a7309},
  author = {Knoop, M.W. and Ricco, P.},
  title = {Dataset underlying the publication: Turbulent skin-friction drag reduction by passively rotating discs},
  publisher = {4TU.ResearchData},
  year = {2026},
  copyright = {CC BY 4.0},}

@article{sillero2013one,
  title={One-point statistics for turbulent wall-bounded flows at {R}eynolds numbers up to $\delta^+ \approx 2000$},
  author={Sillero, J.A. and Jim{\'e}nez, J. and Moser, R.D.},
  journal={Phys. Fluids},
  volume={25},
  number={10},
  year={2013},
  publisher={AIP Publishing},
DOI = {10.1063/1.4823831}
}

@article{lozano2018modeling,
  title={Modeling boundary-layer transition in direct and large-eddy simulations using parabolized stability equations},
  author={Lozano-Dur{\'a}n, A and Hack, MJP and Moin, P},
  journal={Phys Rev. Fluids},
  volume={3},
  number={2},
  pages={023901},
  year={2018},
  publisher={APS},
DOI = {10.1103/PhysRevFluids.3.023901}
}

@article{zaki2013streaks,
  title={From streaks to spots and on to turbulence: exploring the dynamics of boundary layer transition},
  author={Zaki, T.A.},
  journal={Flow Turbul. Combust.},
  volume={91},
  number={3},
  pages={451--473},
  year={2013},
  publisher={Springer},
DOI = {10.1007/s10494-013-9502-8}
}

@article{frohnapfel2024flow,
  title={Flow resistance over heterogeneous roughness made of spanwise-alternating sandpaper strips},
  author={Frohnapfel, B. and von Deyn, L. and Yang, J. and Neuhauser, J. and Stroh, A. and {\"O}rl{\"u}, R. and Gatti, D.},
  journal={J. Fluid Mech.},
  volume={980},
  pages={A31},
  year={2024},
  publisher={Cambridge University Press}
}

@article{knoop2025response,
  title={Response of a turbulent boundary layer to steady, square-wave-type transverse wall-forcing},
  author={Knoop, M.W. and Deshpande, R. and Schrijer, F. F. J. and van Oudheusden, B. W.},
  journal={Phys. Rev. Fluids},
  volume={10},
  number={6},
  pages={064607},
  year={2025},
DOI = {10.1103/PhysRevFluids.10.064607},
  publisher={APS}
}

@book{Cebeci1977Momentum,
       author = {{Cebeci}, T. and {Bradshaw}, P.},
        title = "{Momentum transfer in boundary layers}",
         year = {1977},
        publisher = {McGraw-Hill Book Co},
        adress = {New York}
}

@article{kadivar2021review,
  title={A review on turbulent flow over rough surfaces: Fundamentals and theories},
  author={Kadivar, M. and Tormey, D. and McGranaghan, G.},
  journal={Int. J. Thermofluids},
  volume={10},
  pages={100077},
  year={2021},
  publisher={Elsevier}
}

@article{wangsawijaya2020effect,
  title={The effect of spanwise wavelength of surface heterogeneity on turbulent secondary flows},
  author={Wangsawijaya, D.D. and Baidya, R. and Chung, D. and Marusic, I. and Hutchins, N.},
  journal={J. Fluid Mech.},
  volume={894},
  pages={A7},
  year={2020},
  publisher={Cambridge University Press}
}

@inproceedings{Kempaiah2022Large-scale, 
   author = {Kempaiah, K.U. and Sem, J. and Baars, W.J.},
year = {2022},
   title = {Large-scale flow organization of wall-bounded turbulence over flush-mounted rotating discs},
   booktitle = {12th International Symposium on Turbulence and Shear Flow Phenomena (TSFP12) Osaka, Japan},
   type = {Conference Proceedings}
}

@article{Klewicki2003Laminar,
   author = {Klewicki, J.C. and Hill, R.B.},
   title = {Laminar boundary layer response to rotation of a finite diameter surface patch},
   journal = {Phys. Fluids},
   volume = {15},
   number = {1},
   pages = {101-111},
   ISSN = {1070-6631},
   year = {2003},
   type = {Journal Article}
}

@article{choi1994active,
  title={Active turbulence control for drag reduction in wall-bounded flows},
  author={Choi, H. and Moin, P. and Kim, J.},
  journal={J. Fluid Mech.},
  volume={262},
  pages={75--110},
  year={1994},
  publisher={Cambridge University Press}
}

@article{choi2002near,
  title={Near-wall structure of turbulent boundary layer with spanwise-wall oscillation},
  author={Choi, K.-S.},
  journal={Phys.Fluids},
  volume={14},
  number={7},
  pages={2530--2542},
  year={2002},
  publisher={American Institute of Physics}
}

@article{moin1990direct,
  title={Direct numerical simulation of a three-dimensional turbulent boundary layer},
  author={Moin, P. and Shih, T.-H. and Driver, D. and Mansour, N.N.},
  journal={Phys. Fluids A},
  volume={2},
  number={10},
  pages={1846--1853},
  year={1990},
  publisher={American Institute of Physics}
}

@article{anderson2015numerical,
  title={Numerical and experimental study of mechanisms responsible for turbulent secondary flows in boundary layer flows over spanwise heterogeneous roughness},
  author={Anderson, William and Barros, Julio M and Christensen, Kenneth T and Awasthi, Ankit},
  journal={J. Fluid Mech.},
  volume={768},
  pages={316--347},
  year={2015},
  publisher={Cambridge University Press}
}

@article{vanderwel2015effects,
  title={Effects of spanwise spacing on large-scale secondary flows in rough-wall turbulent boundary layers},
  author={Vanderwel, Christina and Ganapathisubramani, Bharathram},
  journal={J. Fluid Mech.},
  volume={774},
  pages={R2},
  year={2015},
  publisher={Cambridge University Press}
}

@article{bradshaw1987turbulent,
  title={Turbulent secondary flows},
  author={Bradshaw, P.},
  journal={Annu. Rev. Fluid Mech.},
  volume={19},
  pages={53--74},
  year={1987}
}

\end{document}